\documentclass{aa}  

\usepackage{graphicx}
\usepackage{subfigure}
\usepackage[breaklinks,colorlinks,citecolor=blue]{hyperref}
\usepackage[flushleft]{threeparttable}
\usepackage{multirow}
\usepackage{lscape}
\usepackage{longtable}
\usepackage{supertabular}
\usepackage{rotating}
\usepackage{caption}
\usepackage{txfonts}

\usepackage{lscape}

\hypersetup{
  bookmarks=true
  citecolor=blue
  colorlinks   = true, 
  urlcolor     = magenta, 
  linkcolor    = blue, 
  linktoc= none
  backref = false
  pagebackref = false
}

\begin{document}

   \title{The nucleosynthetic history of elements in the Galactic disk:\thanks{Based on observations obtained at the ESO VLT at Paranal Observatory (Observing program 083.D-0871).}}
\subtitle{[X/Fe] - age relations from high-precision spectroscopy}

   \author{L. Spina\inst{\ref{inst1}}
          \and
          J. Mel\'endez\inst{\ref{inst1}}  
          \and
          A. I. Karakas\inst{\ref{inst2}}
          \and
          I. Ram\'irez\inst{\ref{inst3}}
          \and
          T. R. Monroe\inst{\ref{inst4}}
          \and
          M. Asplund\inst{\ref{inst5}}
          \and
          D. Yong\inst{\ref{inst5}}}

  \institute{Universidade de S\~ao Paulo, IAG, Departamento de Astronomia, Rua do Mat\~ao 1226, S\~ao Paulo, 05509-900 SP, Brasil - \email{lspina@usp.br}\label{inst1}
  \and Monash Centre for Astrophysics, School of Physics and Astronomy, Monash University, VIC 3800, Australia\label{inst2}
  \and Department of Astronomy, University of Texas at Austin; 2515 Speedway, Stop C1400, Austin, TX 78712-1205, USA\label{inst3}
  \and Space Telescope Science Institute, 3700 San Martin Drive, Baltimore, MD 21218, USA\label{inst4}
  \and Research School of Astronomy and Astrophysics, The Australian National University, Cotter Road, Canberra, ACT 2611, Australia\label{inst5}
               }

   \date{Received March 18, 2016; Accepted June 15, 2016}

 
  \abstract
   {
The chemical composition of stars is intimately linked to the Galaxy formation and evolution. 
   }
   {
We aim to trace the chemical evolution of the Galactic disk through the inspection of the [X/Fe]-age relations of 24 species from C to Eu.
   }
   {
   Using high-resolution and high-signal-to-noise UVES spectra of nine solar twins, we obtained precise estimates of stellar ages and chemical abundances. These determinations have been integrated with additional accurate age and abundance determinations from recent spectroscopic studies of solar twins existing in the literature, comprising superb abundances with 0.01~dex precision. Based on this data set, we outlined the [X/Fe]-age relations over a time interval of 10~Gyr.
   }
   { 
   We present the [X/Fe] - age relations for 24 elements (C, O, Na, Mg, Al, Si, S, K, Ca, Sc, Ti, V, Cr, Mn, Co, Ni, Cu, Zn, Y, Ba, La, Ce, Nd, and Eu). Each different class of elements showed distinct evolution with time that relies on the different characteristics, rates and timescales of the nucleosynthesis' sites from which they are produced. The $\alpha$-elements are characterised by a [X/Fe] decrement as time goes on. Strikingly, an opposite behaviour is observed for Ca. The iron-peak elements show an early [X/Fe] increase followed by a decrease towards the youngest stars. The [X/Fe] for the $n$-capture elements decrease with age. We also found that both [Mg/Y] and [Al/Y] are precise stellar clocks, with [Al/Y] showing the steepest dependence with age.
   }
   {
   Knowledge of the [X/Fe]-age relations is a gold mine from which we can achieve a great understanding about the processes that governed the formation and evolution of the Milky Way. Through the reverse engineering of these relations we will be able to put strong constraints on the nature of the stellar formation history, the SNe rates, the stellar yields, and the variety of the SNe progenitors.
    }


\authorrunning{L. Spina et al.}
\titlerunning{The nucleosynthetic history of elements in the Galactic disk}

   \maketitle
%

\section{Introduction}
\label{Intr}
``\textit{The principal value of abundance determinations from stellar spectra is the clues they give to the nuclear history of stellar matter and, more generally, of the matter in the whole Galaxy}'' \citep{Cayrel66}.
Fifty years after this pioneering work on stellar chemical abundances, the study of the chemical evolution of our Galaxy is still one of the main research areas in modern astrophysics.

Important hints on the formation and evolution of galaxies are locked into the chemical compositions of stars. All the metals originate from stars that enrich the interstellar medium (ISM) with their own unique pattern of elements depending on their mass and initial metallicity. In fact, each specific element has been produced by different sites of nucleosynthesis that contribute to the chemical evolution of galaxies with different timescales. 

The Type~II supernovae (SNe~II)  are the evolutionary terminus of the most massive stars (M $\gtrsim$ 10 M$_{\sun}$) and are mainly responsible for the production of $\alpha$-elements (e.g., C, O, Mg, Si, S, Ca, Ti). Since the massive stars have very short lifetimes ($\lesssim$10$^{-2}$~Gyr), the SNe~II have polluted the interstellar medium (ISM) since the very early stages of Galaxy's evolution (see \citealt{Matteucci14}). On the other hand, the Type~Ia supernovae (SNe~Ia) explosions are thought to be the final fate of white dwarfs gradually accreting mass from a binary companion (single-degenerate scenario) or the result of the merger of two white dwarfs (double-degenerate scenario); they are important producers of iron-peak elements (e.g., Cr, Mn, Co, Ni). Owing to the relatively low mass of the SNe~Ia progenitors, the timescales in which they operate are typically longer ($\sim$1~Gyr) than those of SNe~II.
However, the SN~Ia play a somewhat ambiguous role in astrophysics due to the large variety of progenitors that are responsible for different SN~Ia subclasses with diverging properties (see \citealt{Hillebrandt13} for a recent review). For instance, among the different SNe~Ia, it has been proposed the existence of a class that operated shortly after ($\sim$100~Myr) the beginning of the stellar formation and that could have played a role in the early Galactic evolution, but they seem a minor fraction ($<$15-20$\%$) of the SNe~Ia \citep{Bonaparte13}.

Last but not the least, all the elements heavier than Zn (Z$>$30) can be produced in different environments; they are commonly dubbed as neutron- ($n$-) capture elements, since they are synthesised in stars through $n$-capture processes: neutrons are fused into nuclei that, if unstable, can $\beta$ decay transforming neutrons into protons. 
There are two types of $n$-capture processes: if the neutron capture timescale is longer than the decay timescale, we will have a slow- ($s$-) process, otherwise a rapid- ($r$-) process will occur. The neutron-capture timescales are largely determined by the neutron number densities ($n_{n}$): the $s$-process occurs when $n_{n}$$\lesssim$10$^{8}$~cm$^{-3}$ \citep{Busso99}, while the $r$-process takes place with much higher densities ($n_{n}$$\sim$10$^{24-28}$~cm$^{-3}$; \citealt{Kratz07}, and references therein). This means that the two processes require very different astrophysical environments: the $s$-process is thought to occur in low- and intermediate-mass stars (i.e., 1-8~M$_{\sun}$) during their AGB phase, while the large $n_{n}$ required for the $r$-process suggest that they likely occur during SN explosions (e.g., \citealt{Arlandini99,Sneden08,Karakas10,Winteler12,Fishlock14,Karakas14}). However, it has also been argued that not all SN explosions can host $r$-process. Namely, only low-mass ($\lesssim$11M$_{\odot}$) SNe are likely sites for the synthesis of $r$-process elements (see \citealt{Cowan04}, and references therein). Nevertheless, other sites have also been hypothesised as responsible of the production of these elements, including colliding neutron stars \citep{Argast04,Korobkin12} and black hole/neutron star mergers \citep{Surman08}. Anyhow, the actual production site(s) of the $r$-process are not known at present \citep{Cowan04,Thielemann11}.

Due to this variety of nucleosynthesis channels, the abundance ratios of two elemental species having different origin and characteristic timescales for release into the ISM, can behave as cosmic clocks by which the formation timescales of various stellar populations (e.g., thin and thick disk) can be determined (e.g., \citealt{Gilmore89,Chiappini97,RecioBlanco14}). Through this information it could also be possible to reveal additional clues on the stellar formation rate, the initial mass function, the mass and metallicity of the SNe progenitors, the occurrence of SNe and AGB stars and their timescales of elemental pollution (e.g., \citealt{Tinsley79,Matteucci86,Yoshii96,Matteucci01,Venn04,Pipino09,Tsujimoto12,Jimenez15,Vincenzo16}). 

In addition, knowledge of the relations between the abundance ratios and age would put strong constraints on all these variables. However, the difficulties in providing precise age determinations of main-sequence stars and the age-metallicity degeneracy of stellar populations prevented for decades from achieving accurate information on the age-abundances dependencies \citep{Edvardsson93}. Hints of these relations have been achieved for $n$-capture elements in field stars \citep{Bensby07,Battistini16}. Additional clues come from the study of the chemical composition of open clusters for which ages are much better constrained than field stars \citep{DOrazi09c, Maiorca11}. These studies revealed an increasing contribution of $s$-process elements from low-mass (1-4 M$_{\sun}$) AGB stars as time goes on. Also, \citet{Magrini09} noted that the [$\alpha$/Fe] ratio is slightly higher for the oldest clusters.

Only recently, a series of spectroscopic studies \citep{Melendez12,Melendez14,Bedell14,Ramirez14,Nissen15,Biazzo15,Spina16} showed that, through accurate equivalent widths (EWs) measurements of high-quality spectra and a strict line-by-line excitation/ionisation balance analysis relative to the solar spectrum, it is possible to achieve chemical abundances at sub-0.01~dex precision and extremely accurate atmospheric parameters and stellar ages.


Using this approach, \citet{Nissen15} (hereafter N15) determined the atmospheric parameters, ages and high-precision abundances of 14 elements from C to Y 
in a sample of 21 solar twins, spanning over an age interval of eight Gyr. This analysis revealed for several elements the existence of tight linear correlations between [X/Fe] and stellar ages. These correlations are positive for most the elements from C to Zn, with the only exception of Ca and Cr, which respectively show flat and a slightly negative correlations. Also Y decreases with increasing age, in agreement with the previous studies on $s$-process elements. On the other hand, N15 noted that the elements Na and Ni seemed to be not well correlated with the stellar ages.
Recently, \citet{Spina16} (hereafter S16) analysed a sample of 14 solar twins and they studied the [X/Fe] vs age relations for all the elements analysed by N15 plus another eight species. 
They confirmed the general behaviour previously found by N15, but for C, O, Na, S, Ni, and Zn they found greater [X/Fe]-age slopes compared to those found by N15. They suggested that this difference may be due to a turnover in the [X/Fe] - age relations occurring at $\sim$6 Gyr. In fact, while the sample studied by N15 covers ages up to 8~Gyr, all the solar twins considered by S16 are younger than 6~Gyr and partially sample the non-linear relations (e.g., see Na in Fig.~\ref{XFe_age}).


In the present paper we aim to provide new key insights on the [X/Fe] vs age relations for 24 species from C to Eu, through the exploitation of high-quality data of solar twins spanning ages from 0.5 to 10 Gyr. In Section~\ref{Data} we describe the spectral analysis and the determination of [X/Fe] ratios and stellar ages. The [X/Fe] vs age relations are discussed in Section~\ref{discussion}. In Section~\ref{conclusions} we present our concluding remarks.

\section{The data set}
\label{Data}

A part of the data set that we employed for our study comes from the observation and spectral analysis of a new study of nine solar twins. The resulting information has been integrated with high-precision [X/Fe] ratios and ages for solar twins taken from the literature.

\subsection{Spectroscopic observations and data analysis}

A sample of nine solar-twins were observed using UVES mounted on the Very Large Telescope during the nights 29-30 August 2009 (program 083.D-0871). The UVES spectrograph was used in the configurations that cover 306-387~nm in the blue and 480-1020~nm in the red chip. Each star was observed with a spectral resolution of R$\sim$65,000 in the blue and 110,000 in the red. The final signal-to-noise ratios (S/N) are within 750 and 1300 pixel$^{-1}$ at 672~nm with a median of $\sim$1000. In addition to the solar-twins, a short exposure of Juno was performed to acquire a solar spectrum with a S/N of 1250 pixel$^{-1}$. The UVES data was reduced by hand using IRAF, as described by \citet{Monroe13} and \citet{TucciMaia15}.

Our method of analysis of the nine solar twin spectra is similar to that described by \citet{Spina16}. Namely, we employed a master list of atomic transitions from \citet{Melendez14} that includes 86 lines of Fe~I, 14 of Fe~II, and $\sim$200 lines of other elements. Using the IRAF \texttt{splot} task and adopting the same approach described by \citet{Bedell14}, we measured in all the solar twin spectra and in the solar spectrum, the EWs of the absorption features listed in the master list. We detected in the spectra absorption features of 24 elements besides iron (C, O, Na, Mg, Al, Si, S, K, Ca, Sc, Ti, V, Cr, Mn, Co, Ni, Cu, Zn, Y, Ba, La, Ce, Nd, Eu). 

We fed the \textit{qoyllur-quipu}\footnote{Qoyllur-quipu or $q^2$ \citep{Ramirez14b} is a Python package that is free and available online at \url{https://github.com/astroChasqui/q2}} code with our EW measurements and it automatically estimated the stellar parameters by performing a line-by-line differential excitation/ionisation balance analysis of the iron EWs relative to the solar spectrum. Namely, the $q^2$ code iteratively searched for the three equilibria (excitation, ionisation, and the trend between the iron abundances and log (EW/$\lambda$)). The iterations were executed with a series of steps starting from a set of initial parameters (i.e., the nominal solar parameters) and arriving at the final set of parameters that simultaneously fulfil the equilibria. Further details on this procedure can be found in \citet{Ramirez14b}. For our analysis we adopted the Kurucz (ATLAS9) grid of model atmospheres \citep{Castelli04} and we assumed the following solar parameters: $T_{eff}$$=$5777~K, log~g$=$4.44~dex, [Fe/H]$=$0.00~dex and $\xi=$1.00~km/s. The errors associated with the stellar parameters are evaluated by the code following the procedure described in \citet{Epstein10} and \citet{Bensby14}. Since each atmospheric parameter is dependent on the others, this approach takes into account this dependence by propagating the error associated with the fulfilment of the three equilibrium conditions in every single parameter. The resulting stellar parameters and their uncertainties are reported in the first columns on Table~\ref{parameters}.

Once the stellar parameters and the relative uncertainties have been determined for each star, $q^2$ automatically employs the appropriate atmospheric model for the local thermodynamic equilibrium (LTE) calculation of the chemical abundances through the 2014 version of MOOG \citep{Sneden73}. All the elemental abundances are normalised relative to the solar values on a line-by-line basis. The $q^2$ code calculates the error budget associated with the abundances [X/H] taking into account the observational error due to the line-to-line scatter from the EW measurements (standard error), and the errors in the atmospheric parameters. When, as for K, Ce, La, and Eu, only one line is detected, the observational error was estimated by repeating the EW measurement five times with different assumptions on the continuum setting, adopting as error the standard deviation.

The $q^2$ code also calculates the non-LTE (NLTE) corrections for the oxygen abundances as a function of the stellar parameters assuming the NLTE corrections computed by \citet{Ramirez07}. 
As shown by \citet{Amarsi16}, for Sun-like stars these corrections are in good agreement with the more recent grids of NLTE corrections provided by \citet{Fabbian09} and \citet{Amarsi16}. Nevertheless, we performed a quantitative test in order to evaluate how much our final oxygen abundances are affected by the choice of different grids of NLTE corrections. For this test we considered the two stars for which we expect the most extreme NLTE corrections to be applied: HIP73815, which has the most discrepant log~g from the solar one, and HIP77883, that has the most discrepant T$_{eff}$ from the solar value. We found that the differential corrections in oxygen abundance between the grid of Amarsi et al. and that of Ram\'irez et al. is 0.003~dex for HIP73815 and 0.002~dex for HIP77883\footnote{The grid of Ram\'irez et al. establishes a oxygen NLTE corrections of $-$0.022~dex for HIP73815 and 0.000~dex for HIP77883, while the grid of Amarsi et al. predicts corrections of $-$0.025 and +0.002~dex, respectively.}. These quantities are much smaller than the uncertainties that we obtained for the oxygen abundances (i.e. $\geq$0.007~dex). For practical reasons we keep the oxygen abundances corrected for the Ramirez's grid, since that grid is already implemented in the $q^2$ code.

In addition to this test, we verified on the same two stars if NLTE corrections for additional elements should be taken into account. Based on the grid provided by \citet{Bergemann12} we applied the differential NLTE corrections on 28 iron lines, that embody $\sim$30$\%$ of all the iron lines measured in the spectra of these stars. All other iron lines with EW measurements are not included in that grid. We found that differential NLTE effects are negligible for iron, since it would account for a [Fe/H] increment of 0.002~dex in HIP73815, while the iron abundance of HIP77883 remained unaltered. We repeated the experiment for Na using the grid provided by \citet{Lind11}. For the star HIP73815, the NLTE Na abundance differs from that obtained through MOOG by 0.003~dex. As for iron, no differences have been seen for HIP77883. Then we employed the grid described in \citet{Melendez16} to derive these differential corrections also for Mg and Ca. Also for these two elements, the NLTE effects are negligible, since are less than 0.005 and 0.004~dex for Mg and Ca, respectively. These experiments$\footnote{The NLTE corrections for Fa and Na have been calculated through the interactive database INSPECT v1.0 available online at \url{www.inspect-stars.com}. For Mg and Ca we adopted an iterpolator written for IDL.}$ show that the differential NLTE corrections for Fe, Na, Mg, and Ca are small relatively to uncertainties that we obtained for the abundances of these elements. As it has been shown by other authors \citep{Melendez12,Melendez16,Nissen15}, differential NLTE effects in solar twins are negligible for all the other elements that we detected. According to this, the oxygen is the only element that has been corrected for differential NLTE effects.

In addition, through the \textit{blends} driver in the MOOG code, we took into account the hyperfine/isotopic splitting affecting the following elements Sc, V, Mn, Co, Cu, Y, Ba, La, and Eu. We assumed the HSF line list from the Kurucz database\footnote{The Kurucz database of atomic transitions is free and available at \url{http://kurucz.harvard.edu/linelists.html}.}.  

Given the atmospheric parameters and their uncertainties, the $q^2$ code also calculates the age and mass probability distributions for each star according to the Yonsei-Yale isochrone set \citep{Yi01,Kim02}. Details on this procedure are given by \citet{Ramirez13,Ramirez14}.

\begin{table*}
\begin{center}
\caption{\label{parameters} Stellar parameters.}
\begin{tabular}{c|cccc|cc} 
\hline\hline 
Star & $T_{eff}$ & log g & [Fe/H] & $\xi$ & Age & Mass \\
 & [K] & [dex] & [dex] & [km s$^{-1}$] & [Gyr] & [$M_{\sun}$] \\ \hline
18Sco & 5818$\pm$3 & 4.457$\pm$0.010 & 0.054$\pm$0.004 & 1.01$\pm$0.01 & 3.0$\pm$0.4 & 1.040$\pm$0.002 \\
HIP56948 & 5793$\pm$6 & 4.473$\pm$0.022 & 0.019$\pm$0.005 & 1.01$\pm$0.01 & 2.6$\pm$1.0 & 1.024$\pm$0.008 \\
HIP102152 & 5724$\pm$3 & 4.357$\pm$0.009 & $-$0.013$\pm$0.003 & 0.99$\pm$0.01 & 8.0$\pm$0.3 & 0.971$\pm$0.003 \\
HD20630 & 5747$\pm$8 & 4.510$\pm$0.016 & 0.065$\pm$0.007 & 1.12$\pm$0.01 & 1.4$\pm$0.6 & 1.025$\pm$0.006 \\
HD202628 & 5836$\pm$7 & 4.503$\pm$0.013 & 0.009$\pm$0.005 & 1.04$\pm$0.01 & 0.8$\pm$0.6 & 1.047$\pm$0.008  \\
HIP30502 & 5741$\pm$3 & 4.435$\pm$0.008 & $-$0.053$\pm$0.003 & 0.98$\pm$0.01 & 5.7$\pm$0.4 & 0.969$\pm$0.003 \\
HIP73815 & 5796$\pm$3 & 4.320$\pm$0.008 & 0.028$\pm$0.003 & 1.07$\pm$0.01 & 7.1$\pm$0.2 & 1.012$\pm$0.004 \\
HIP77883 & 5702$\pm$3 & 4.378$\pm$0.011 & 0.023$\pm$0.003 & 0.97$\pm$0.01 & 7.5$\pm$0.4 & 0.971$\pm$0.003 \\
HIP89650 & 5849$\pm$3 & 4.403$\pm$0.011 & $-$0.013$\pm$0.003 & 1.07$\pm$0.01 & 5.1$\pm$0.5 & 1.017$\pm$0.005 \\ \hline\hline
\end{tabular}
\end{center}
\end{table*}

The analysis described in this Section allowed us to determine the atmospheric parameters, ages and masses for the nine solar twins observed through the UVES spectrograph. These values are listed in Table~\ref{parameters}. The sample fully covers the age range of thin disk stars, from 0.8 to 8.0~Gyr. The [Fe/H] values are within $-$0.053 and 0.065~dex. The typical precisions that we reached are $\sigma$($T_{eff}$)$=$4~K, $\sigma$(log~g)$=$0.012~dex, $\sigma$([Fe/H])$=$0.004~dex, $\sigma$($\xi$)$=$0.01~km/s, $\sigma$(Age)$=$0.5~Gyr and $\sigma$(Mass)$=$0.005~M$_{\sun}$. All the chemical abundances are listed in Table~\ref{XFe_ratios}.

Recently, \citet{Ramirez12} analysed the kinematics of seven solar twins of our sample establishing that they belong to the thin disk population. Assuming the parallaxes and proper motions from \citet{vanLeeuwen07}, and the radial velocities from \citet{Soubiran13}, we determined that HD20630 and HIP73815, the two stars of our sample not analysed by \citet{Ramirez12}, have a probability $\geq$0.97 of belonging to the thin disk population according to the membership criteria by \citet{Reddy06}.

\subsection{The data set from the literature}
We integrated the dataset achieved from our UVES observations with the [X/Fe] and age determinations provided by N15 and S16. The former ensured a dataset of 18 thin disk stars and 3 high-$\alpha$-metal-rich stars, which are a population first identified by \citet{Adibekyan11}. As suggested by \citet{Haywood13}, this population is part of the thick disk and is characterised by having higher [$<$$\alpha$$>$/Fe] values ($\sim$0.1 dex) with respect to the normal thin disk stars. For simplicity, hereafter we will refer to the high-$\alpha$-metal-rich stars as ``thick disk stars''. For the star 18Sco, which has been analysed both in the present work and by N15, we adopt the values determined here. We note that the agreement between our [X/Fe] ratios and those of N15 is within one sigma with very high confidence given that the mean difference $\Delta$[X/Fe] (N15 $-$ this work) $=$ 0.006~dex ($\sigma$=0.006~dex) and that only [O/Fe] (out of 12 elements) have a difference that is marginally over the 1-sigma error.

In addition we included the dataset from S16, who analysed 13 thin disk and one thick disk star, determining ages and [X/Fe] ratios. One star, HD45184, has been analysed by both S16 and N15. As for 18Sco, a high level of consistency resulted from the atmospheric parameters and abundances values determined for this star by N15 and S16, indicating that the data set provided by our group is homogeneous with that of Nissen. For HD45184 we assumed, when available, the N15 values that are based on higher-quality spectra, otherwise we used the S16 determinations.

We also included in our dataset the age and abundances determinations provided by \citet{YanaGalarza16a} for HIP~100963 and the abundances determined by \citet{TucciMaia14} for 16Cyg~A and 16Cyg~B. For 16Cyg~A/B, we corrected the oxygen abundances for the NLTE effects through the same procedure described in Section~\ref{Data}. Since \citet{TucciMaia14} did not analyse the $n$-capture elements, the [X/Fe] ratios of these latter have been taken from \citet{Ramirez11}, as well as the age determination of this system. 

Our final dataset includes [X/Fe] ratios and ages for 41 thin disk and four thick disk stars.

\section{[X/Fe] vs age relations}
\label{discussion}

As we mentioned in the introduction, chemical properties of stars provide extremely important insights on the mechanisms responsible for the Galaxy formation and evolution. In this section we discuss, first, the [X/Fe] vs age relations for the elements from C to Zn, then those for the $n$-capture elements (i.e., from Y to Eu).

\subsection{The elements from C to Zn}
\begin{figure*}
\centering
\includegraphics[width=16.5cm]{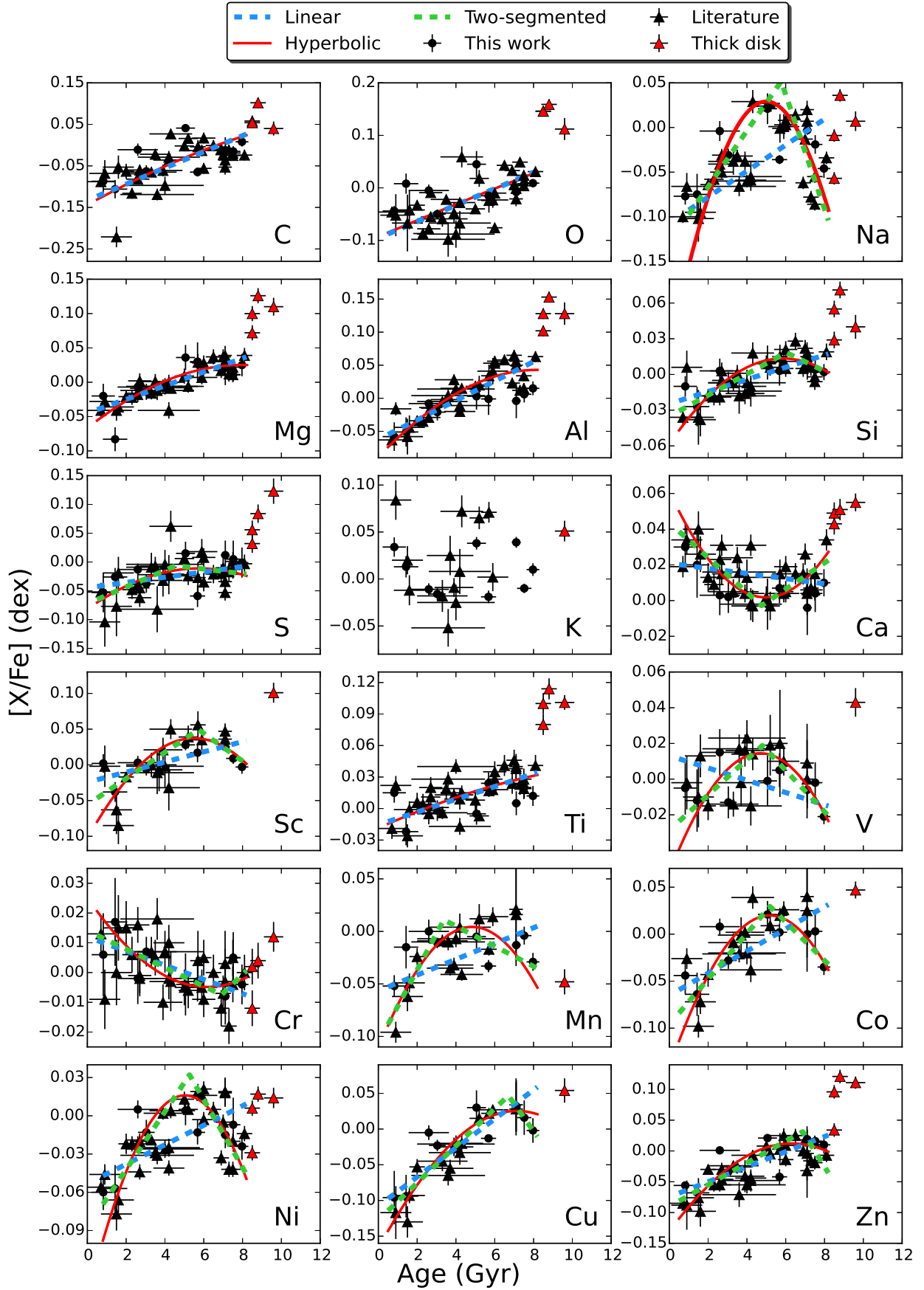} \caption{[X/Fe] ratios of the elements with Z$\leq$30 as a function of stellar ages for the nine solar twins analysed here (circles). We also plotted the measurements taken from the literature as triangles. The red triangles represent the thick disk stars. The [X/Fe]-age distributions have been fitted by linear, hyperbolic and two-segmented line functions showed as blue dashed, red solid and green dashed lines, respectively. The parameters related to these functions are listed in Table~\ref{age_fit}. Note that the y-axes ranges are different from element to element, in order to properly visualise our entire data set.}
\label{XFe_age}
\end{figure*}

The [X/Fe] vs age relations for the elements with Z$\leq$30 are plotted in Fig.~\ref{XFe_age}. We note that the [X/Fe] ratios of most of the species are highly correlated with the stellar ages and that each element is characterised by its own relation. Fig.~\ref{XFe_age} also shows that the thick disk stars do not necessarily follow the main distributions traced by the thin disk population. However, for some elements, like Ca and Cr, they appear to follow the thin disk distribution.
The difference between thin and thick disk stars is clearly visible in the $\alpha$-elements of which this thick disk population is systematically overabundant (as noted also by \citealt{Adibekyan11,Haywood13}), but also in some iron-peak elements (such as V, Co, and Ni), and in Na, Cu and Zn. For this reason we treat the thick disk stars separately from the sample of thin disk stars. As we can see, the boundary between the thin and thick disk populations is located at $\sim$8 Gyr ago. This age is in perfect agreement with the formation age of the thin disk based on the G-dwarf metallicity distribution for stars in the solar vicinity (e.g., \citealt{Chiappini97}).

From 8~Gyr towards younger ages, the chemical content of the thin disk highly evolved through time, with each specie undergoing different characteristic types of evolution. Each peculiar behaviour must reflect the variety of the yields and timescales due to different sites of synthesis. Most of the [X/Fe]-age distributions seem nonlinear and, in some cases, also non-monotonic. Only potassium \text{does not exhibit} a definite dependency with age, probably because its abundances are only based on the saturated line at 7698.97~$\AA$; and also some weak telluric contamination may be present.

\subsubsection{Fitting of the [X/Fe]-age relations}
For all the elements, except potassium, we proceed to fit the [X/Fe] vs age relations traced by the 41 thin disk stars assuming three different models and using the orthogonal distance regression method \citep{Boggs90}, which takes into account the uncertainties on both [X/Fe] and ages\footnote{For the data fitting we used the python scipy.odr package (\url{http://docs.scipy.org/doc/scipy/reference/odr.html}).}. First, we performed the fit employing a straight line function, defined as follows:
\[
\left[ \frac{X}{Fe} \right]=a+b\times \text{Age[Gyr]}.
\]
This very simple model has only two parameters, and it has been assumed also by N15 and S16 for their data fitting. The final parameters and their uncertainties are listed in the second and third columns of Table~\ref{age_fit} and plotted in Fig.~\ref{XFe_age} as blue dashed lines. We note that these slopes are in agreement with those of N15 within their uncertainties for all the elements, except for carbon. In the first three columns of Table~\ref{stat_fit}, we reported for each element the average of the residuals ($\sigma_{[X/Fe]}$) and the reduced chi-square ($\chi^{2}_{red}$) obtained from the linear fitting. From these values we noted that some elements, such as Mg and Cr, have $\sigma_{[X/Fe]}$ values that are smaller than typical uncertainties on their [X/Fe] ratios and that have $\chi^{2}_{red}$$\sim$1. On the other hand, the straight line model completely fails in fitting the [X/Fe]-age relations of Na and Ni. Similar differences in the fit goodness have been found also by N15 when comparing the same elements. As mentioned in the introduction, S16 attributed this large dissimilarity to turnovers that may be present in the [X/Fe]-age distributions.

In order to verify if these relations are linear or if they have turnovers as suggested by S16, we repeated the fitting procedure using a model based on a general hyperbolic function
\[
\left[ \frac{X}{Fe} \right]=a+\sqrt{ \frac{\left(\text{Age[Gyr]}-k\right)^2}{b^2}+1}.
\]
This model has three parameters and it would provide rough, but adequate estimations of the ages correspondent to the turnovers or stationary points (hereafter, knees) of the [X/Fe] vs age distributions (the ``k'' parameter in the equation above). The resulting parameters are reported in the fourth, fifth and sixth columns of Table~\ref{age_fit} and plotted in Fig.~\ref{XFe_age} as solid red lines. The parameters of the hyperbolic functions are satisfactorily constrained with the only exceptions of elements like C, O, Mg, Al, and Ti, whose [X/Fe] ratios show a quasi-linear dependence with age. For these elements we report in Table~\ref{age_fit} a lower limit for the k parameter at 7~Gyr. Comparing the $\chi^{2}_{red}$ of the linear fit to those of hyperbolic fit, reported in Table~\ref{stat_fit}, we note that the hyperbolic model resulted in lower $\chi^{2}_{red}$ values for the majority of the elements, except for C, O, Al, and Ti.

The third model used to fit the data set is based on a two-segmented function, defined as follows:
\[ 
\left[ \frac{X}{Fe} \right] =
  \begin{cases}
    a+b\times \text{Age[Gyr]}       & \quad \text{if } \text{Age[Gyr]} \leq k\\
    a+(b-c)\times k+c\times \text{Age[Gyr]}       & \quad \text{if } \text{Age[Gyr]} > k\\
  \end{cases}.
\]
We did not perform this fit on the elements whose parameters have not been strongly constrained through the hyperbolic model (i.e., C, O, Mg, Al, and Ti), since the fitting procedure would result in functions very similar to straight lines. The parameters are reported in the last four columns of Table~\ref{age_fit} and the functions plotted in Fig.~\ref{XFe_age} as dashed green lines. The two-segmented line model is the most complex of the three models adopted for the fitting procedure (it has four parameters), but it allows us to verify if we can recover knee-ages similar to those obtained through the hyperbolic model. The values of the ``k'' parameter obtained from the two models are consistent within the uncertainties, except for Na and Mn.

In order to evaluate which of the three models provides the best fit of the [X/Fe]-age relations, we conducted a statistical comparison of the results based 
on the F ratio \citep{Fisher25}.
 This quantity associates the relative decrease in $\chi^{2}$ due to the use of two different models in the fitting procedure to the relative decrease in degrees of freedom. Namely, the F ratio is defined as 
 \[
\text{F}=\frac{(\chi^{2}_{1}-\chi^{2}_{2})/\chi^{2}_{2}}{(df_{1}-df_{2})/df_{2}},
\]
where $\chi^{2}_{1}$ and $df_{1}$ are the chi-square and the degree of freedom of the simplest model. Thus, when F$>$1 the relative decrease of $\chi^{2}$ is greater than the relative decrease of degrees of freedom. There are two possible explanations for this case: either the more complex model is the best or the simpler model is correct, but random scatter makes the more complex model fitting better. In this instance, the knowledge of P values allows us to identify which is the correct explanation among the two, since it represents the probability that a random scatter is dominating the fitting \citep{Pearson00,Fisher25}. In the last four columns of Table~\ref{stat_fit} we reported the F ratios and the P values obtained by comparing the linear model with the hyperbolic model (columns eight and nine) and the linear model with the two-segmented line model (columns ten and eleven). We did not calculate the P value when F$<$1. These quantities clearly indicate that the two more complex models (i.e., the hyperbolas and the two-segmented lines) better reproduce the data set, except for C, O, Al, and Ti, which have F ratios $<$1. Among the 13 elements with F$_{Lin-Hyp}$$>$1, all the species have P$_{Lin-Hyp}$$<$0.07 and only three have P$_{Lin-Hyp}$$>$0.01 (i.e., Mg, S, and Cr). Similar results have been found when comparing the straight line model with the two-segmented line model. The outcomes of this statistical comparison between the three models demonstrates that indeed most of the [X/Fe]-age relations analysed in this paper are characterised by a knee located at different ages within 4 and 7 Gyr. As we will detail in Section~\ref{Evolution}, the knee ages depend on the nature of the specific element. In addition, the fact that F$_{Lin-Hyp}$ is always greater than F$_{Lin-2SL}$ indicates that the hyperbolic model is the function that better fits the data among the two non-linear models. For this reason, hereafter we will discuss only the parameterisation of the [X/Fe]-age relations defined through the hyperbolic model.



\begin{table*}
\tiny
\begin{center}
\begin{threeparttable}
\caption{\label{age_fit} Results of the fitting of the elements with Z$\leq$30}
\begin{tabular}{c||cc|ccc|cccc} 
\hline\hline
 & \multicolumn{2}{c|}{Linear}  & \multicolumn{3}{c|}{Hyperbolic} & \multicolumn{4}{c}{Two-Segmented line}\\ \hline
Element & a & b & k & a & b & k & a & b & c \\
 & [dex] & [dex Gyr$^{-1}$] & [Gyr] & [dex] & [Gyr dex$^{-1}$] & [Gyr] & [dex] & [dex Gyr$^{-1}$] & [dex Gyr$^{-1}$] \\ \hline
C & $-$0.133$\pm$0.018 &  0.030$\pm$0.003 & $\gtrsim$7$^{*}$ & $\sim$1.085 & $\sim$22.8 & --- & --- & --- & ---\\
O & $-$0.095$\pm$0.013 & 0.016$\pm$0.002 & $\gtrsim$7$^{*}$ & $\sim$1.356 & $\sim$45.8  & --- & --- & --- & --- \\
Na & $-$0.107$\pm$0.019 & 0.015$\pm$0.003 & 5.0$\pm$0.1 & 1.029$\pm$0.005 & $-$6.3$\pm$0.4  & 5.8$\pm$0.3 & $-$0.129$\pm$0.021 & 0.0314$\pm$0.0076 & $-$0.064$\pm$0.017\\
Mg & $-$0.045$\pm$0.005 & 0.0100$\pm$0.0009 & $\gtrsim$7$^{*}$ & $\sim$1.025 & $\sim$18.2  & --- & --- & --- & ---\\
Al & $-$0.062$\pm$0.008 & 0.0147$\pm$0.0015 & $\gtrsim$7$^{*}$ & $\sim$1.043 & $\sim$15.5  & --- & --- & --- & ---\\
Si & $-$0.025$\pm$0.004 & 0.0051$\pm$0.0008 & 5.9$\pm$0.3 & 1.013$\pm$0.002 & 15.2$\pm$1.8  & 6.0$\pm$0.4 & $-$0.035$\pm$0.005 & 0.0089$\pm$0.0014 & $-$0.008$\pm$0.004\\
S & $-$0.045$\pm$0.009 & 0.0046$\pm$0.0017 & 5.8$\pm$0.6 & 0.989$\pm$0.005 & 15.1$\pm$3.6  & 4.6$\pm$0.9 & $-$0.072$\pm$0.017 & 0.0145$\pm$0.0062 & $-$0.004$\pm$0.005\\
Ca & 0.021$\pm$0.004 & $-$0.0014$\pm$0.0008 & 5.0$\pm$0.2 & $-$0.998$\pm$0.003 & 14.2$\pm$1.4 & 4.8$\pm$0.5 & 0.044$\pm$0.007 & $-$0.0098$\pm$0.0025 & 0.007$\pm$0.003\\
Sc & $-$0.024$\pm$0.010 & 0.0071$\pm$0.0018 & 5.5$\pm$0.2 & 1.037$\pm$0.005 & 10.0$\pm$1.0  & 5.7$\pm$0.5 & $-$0.056$\pm$0.013 & 0.0184$\pm$0.0037 & $-$0.019$\pm$0.009\\
Ti & $-$0.016$\pm$0.005 & 0.0061$\pm$0.0011 & $\gtrsim$7$^{*}$ & $\sim$1.045 & $\sim$41.2 & --- & --- & --- & ---\\
V & 0.013$\pm$0.007 & $-$0.0034$\pm$0.0012 & 4.7$\pm$0.2 & 1.014$\pm$0.003 & 12.6$\pm$1.2 & 4.8$\pm$0.5 & $-$0.029$\pm$0.009 & 0.0099$\pm$0.0031 & $-$0.012$\pm$0.003\\
Cr & 0.012$\pm$0.003 & $-$0.0024$\pm$0.0005 & 6.2$\pm$0.7 & $-$1.005$\pm$0.002 & 25.0$\pm$5.1 & 6.9$\pm$0.5 & 0.014$\pm$0.003 & $-$0.0031$\pm$0.0007 & 0.006$\pm$0.006\\
Mn & $-$0.057$\pm$0.010 & 0.0075$\pm$0.0019 & 4.8$\pm$0.2 & 1.004$\pm$0.008 & 9.8$\pm$1.5 & 3.4$\pm$0.7 & $-$0.106$\pm$0.041 & 0.0347$\pm$0.0187 & $-$0.010$\pm$0.005\\
Co & $-$0.065$\pm$0.015 & 0.012$\pm$0.003 & 5.2$\pm$0.2 & 1.020$\pm$0.004 & 8.8$\pm$0.7 & 5.2$\pm$0.3 & $-$0.096$\pm$0.014 & 0.0242$\pm$0.0040 & $-$0.020$\pm$0.006\\
Ni & $-$0.053$\pm$0.010 & 0.0076$\pm$0.0017 & 5.0$\pm$0.1 & 1.016$\pm$0.004 & 8.6$\pm$0.6 & 5.3$\pm$0.4 & $-$0.088$\pm$0.015 & 0.0228$\pm$0.0052 & $-$0.027$\pm$0.006\\
Cu & $-$0.108$\pm$0.013 & 0.020$\pm$0.003 & 7.1$\pm$0.8 & 1.025$\pm$0.008 & 11.0$\pm$1.8 & 6.7$\pm$0.5 & $-$0.127$\pm$0.014 & 0.0260$\pm$0.0033 & $-$0.039$\pm$0.037\\
Zn & $-$0.075$\pm$0.007 & 0.0122$\pm$0.0014 & 6.3$\pm$0.4 & 1.012$\pm$0.004 & 11.4$\pm$1.4 & 6.9$\pm$0.3 & $-$0.090$\pm$0.009 & 0.0177$\pm$0.0021 & $-$0.051$\pm$0.035\\
\hline\hline 
\end{tabular}
\begin{tablenotes}
      \tiny
      \item *) The values used for the functions plotted in Fig.~\ref{XFe_age} are 16.3, 48.0, 8.0, 8.2, and 15.0 Gyr for C, O Mg, Al, and Ti, respectively.
    \end{tablenotes}
\end{threeparttable}
\end{center}
\end{table*}

\begin{table*}
\tiny
\begin{center}
\begin{threeparttable}
\caption{\label{stat_fit} Statistical tests resulting from the fitting of the elements with Z$\leq$30}
\begin{tabular}{c|ccc|cc|cc||cc|cc} 
\hline\hline
 & \multicolumn{3}{c|}{Linear} & \multicolumn{2}{c|}{Hyperbolic} & \multicolumn{2}{c||}{Two-Segmented line} & \multicolumn{4}{c}{Statistical model comparison} \\ \hline
Element &  $\sigma_{[X/Fe]}$ & $\chi^{2}_{red}$ & df$^{*}$ & $\sigma_{[X/Fe]}$ & $\chi^{2}_{red}$ & $\sigma_{[X/Fe]}$ & $\chi^{2}_{red}$ & F$_{Lin-Hyp}$ & P$_{Lin-Hyp}$ & F$_{Lin-2SL}$ & P$_{Lin-2SL}$\\
 & [dex] & & & [dex] & & [dex] & & & \\ \hline
C & 0.033 & 4.86 & 39 & 0.033 & 5.04 & --- & ---& $-$0.37 & --- & --- & --- \\
O & 0.028 & 5.19 & 39 & 0.028 & 5.37 & --- & --- & $-$0.27 & --- & --- & ---\\
Na & 0.030 & 12.46 & 39 & 0.033 & 1.89 & 0.028 & 1.98 & 218.94 & 1.11$^{-16}$ & 106.20 & 4.44$^{-16}$\\
Mg & 0.011 & 1.04 & 39 & 0.011 & 0.91 & --- & --- & 6.22 & 1.71$^{-2}$  & --- & ---\\
Al & 0.017 & 2.56 & 39 & 0.015 & 2.61 & --- & --- & 0.21 & ---  & --- & ---\\
Si & 0.010 & 1.60 & 39 & 0.010 & 1.16 & 0.009 & 1.14 & 15.8 & 3.03$^{-4}$ & 8.54 & 8.95$^{-4}$ \\
S & 0.020 & 1.51 & 38 &  0.021 & 1.40 & 0.020 & 1.33 & 3.97 & 5.36$^{-2}$  & 3.33 & 4.70$^{-2}$\\
Ca & 0.009 & 2.23 & 39 & 0.009 & 1.28 & 0.008 & 1.34 & 30.01 & 2.89$^{-6}$ & 14.19 & 2.66$^{-5}$ \\
Sc & 0.021 & 2.67 & 23 & 0.021 & 1.20 & 0.019 & 1.33 & 29.19 & 2.00$^{-5}$ & 13.37 & 1.80$^{-4}$\\
Ti & 0.012 & 2.13 & 39 & 0.012 & 2.21 & --- & --- & $-$0.40 & --- & --- & ---\\
V & 0.014 & 1.92 & 23 & 0.011 & 0.85 & 0.011 & 0.82 & 29.84 & 1.73$^{-5}$  & 15.12 & 8.56$^{-5}$\\
Cr & 0.006 & 1.03 & 39 & 0.006 & 0.96 & 0.006 & 0.93 & 3.57 & 6.64$^{-2}$ & 2.87 & 6.94$^{-2}$\\
Mn & 0.019 & 4.14 & 23 & 0.019 & 2.26 & 0.020 & 2.71 & 20.17 & 1.82$^{-4}$ & 8.08 & 2.50$^{-3}$\\
Co & 0.021 & 4.39 & 23 & 0.020 & 0.96 & 0.020 & 1.06 & 83.68 & 5.95$^{-9}$ & 39.36 & 7.87$^{-8}$\\
Ni & 0.018 & 7.51 & 39 & 0.019 & 1.61 & 0.017 & 1.77 & 143.85 & 1.73$^{-14}$ & 68.77 & 3.44$^{-13}$\\
Cu & 0.019 & 1.62 & 23 & 0.017 & 1.21 & 0.013 & 1.05 & 8.94 & 6.74$^{-3}$ & 6.03 & 8.53$^{-3}$ \\
Zn & 0.019 & 2.34 & 39 & 0.018 & 1.80 & 0.019 & 1.55 & 12.70 & 1.01$^{-3}$ & 9.17 & 5.88$^{-4}$ \\

\hline\hline 
\end{tabular}
\begin{tablenotes}
      \tiny
      \item *) The degrees of freedom (df) are listed only for the linear fitting (two parameters). These related to the hyperbolic (three parameters) and two-segmented line (four parameters) fitting are immediately deducible.
    \end{tablenotes}
\end{threeparttable}
\end{center}
\end{table*}

\subsubsection{Evolution of the [X/Fe] ratios}
\label{Evolution}

The hyperbolic functions overplotted in Fig.~\ref{XFe_age} and parametrised in Table~\ref{age_fit} highlight the peculiarities of each element in the [X/Fe] vs age relations that reflect their various histories of nucleosynthesis. We note that most of the elements are characterised by having an early [X/Fe] increment as times goes on, that is followed by a decrement for the youngest stars (e.g., see Na and Ni). The age of the knee can vary significantly from 4~Gyr to older ages. Some other species (i.e., O, Mg, Al, and Ti) do not show the early increment, however their distributions flatten at 6-8~Gyr. Only Ca and Cr present an early [X/Fe] decrement with time and a subsequent rise.

\begin{figure}
\centering
\includegraphics[width=9cm]{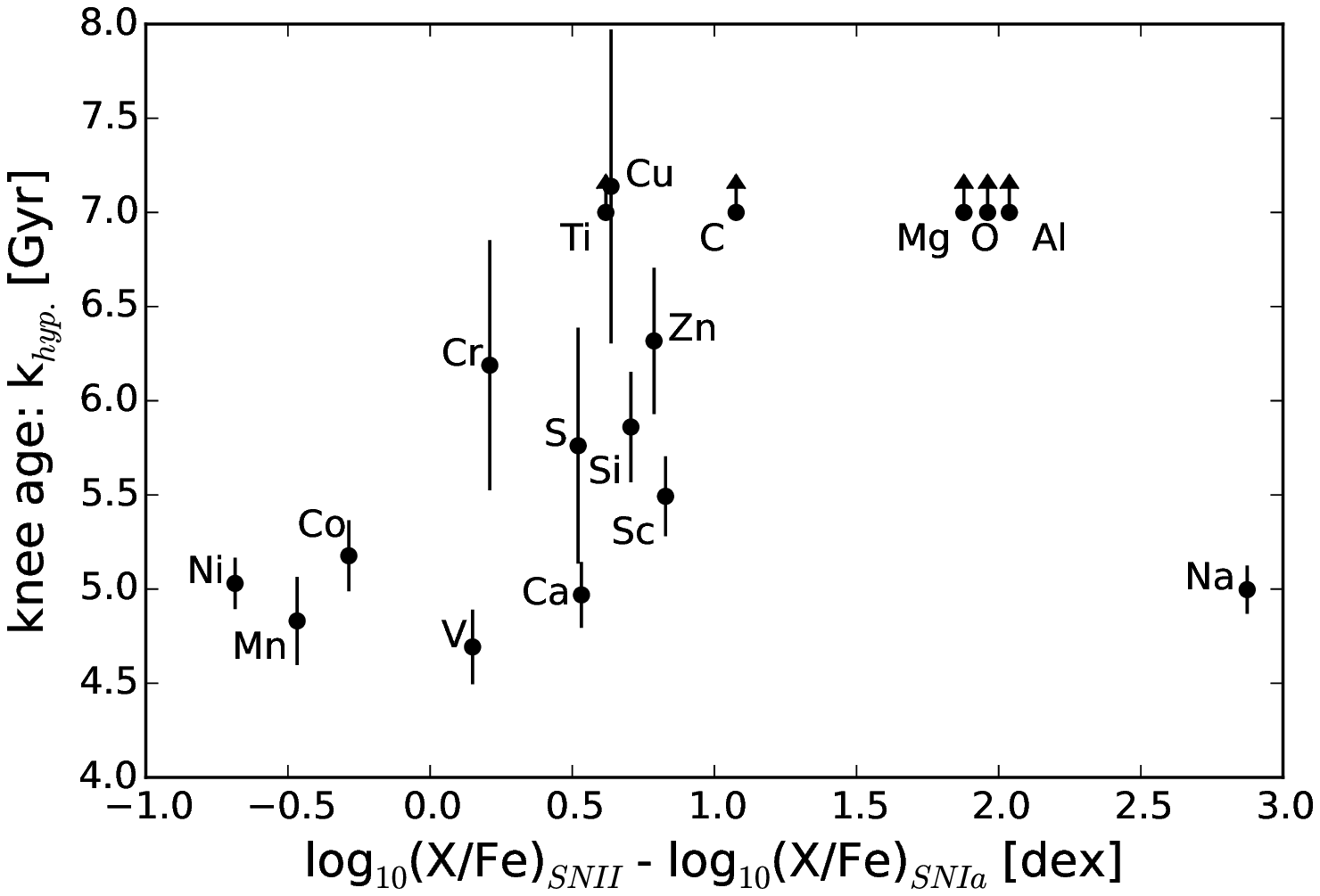} \caption{Ages for the location of the knee in the [X/Fe]-age relations (i.e., the k parameters from the hyperbolic fitting in Table~\ref{age_fit}) as a function of logarithmic ratios between the yields of SNe~II and SNe~Ia \citep{Nomoto97a,Nomoto97b}.}
\label{yields}
\end{figure}

Assuming the yields from SNe II and SNe Ia (W7 model) listed in \citet{Nomoto97a,Nomoto97b}, we plot in Fig~\ref{yields} the knee-ages of the various elements as a function of the ratios between the yields of the two types of SNe. 
Interestingly, most of the elements that are mainly produced by SNe~II  (i.e, the $\alpha$-elements) are those with a knee-age$\gtrsim$6~Gyr. Contrarily, all the elements yielded primarily by SNe~Ia (i.e., the iron-peak elements) have knee-ages $<$5.5~Gyr.

As mentioned in the introduction, the $\alpha$-elements are produced primarily by SNe II events due to high-mass stars of negligible lifetimes in comparison to galaxy evolution timescales. On the other hand, iron is spread into the ISM by both SNe~II and SNe~Ia, where SNe~Ia operate on longer timescales. Therefore, the oldest stars of the thin disk are those formed shortly after the ISM has been enriched by SNe II, thus they should show higher [$\alpha$/Fe] ratios. Besides, the stars that formed sometime after the SNe Ia pollution will have higher iron abundances and lower [$\alpha$/Fe] ratios. As we can observe from Fig. \ref{XFe_age}, the decrease of the [$\alpha$/Fe] ratios began at very old ages, at $\sim$7~Gyr or even earlier. This would suggest that the enrichment from SNe~Ia started to be significant within $\sim$1~Gyr after the formation of the thin disk.

The unique element that does not fit the explanation is Na, that, like the $\alpha$-elements, is mainly produced by SNe~II (see Fig.~\ref{yields}), but its knee-age value is similar to that typical of iron-peak elements ($\sim$5~Gyr). This peculiar behaviour is somewhat expected, since the synthesis of Na is controlled by the neutron excess \citep{Timmes95}, meaning that the Na mass produced in SNe is highly dependent from the stellar metallicity (see Fig. 5 in \citealt{Kobayashi06}). Thus, the first generation of stars, those with the lowest metallicity, did not deliver to the ISM as much Na as the later SNe progenitor did. This would explain why the thin disk underwent a gradual [Na/Fe] increase during its first 2-3~Gyr, with a knee-age that is somewhat delayed in time with respect to that of other $\alpha$-elements, like Si or S. Analogously to Na, Sc is also mainly synthesised by SNe~II, with a high dependency from the metallicity of the progenitor \citep{Kobayashi06}: the mass of Sc produced by stars with solar metallicity (Z$=$0.02) is more than ten times the mass produced by stars with Z=0. Also for this element we observe a knee that is slightly delayed in comparison with other $\alpha$-elements. 

Since the enrichment from massive stars lasts only while the stellar formation is active, the rate of SNe~II is expected to decrease after the first few Gyrs from the formation of the Galactic disk \citep{Gibson03}, quenching their contribution to the pollution of the ISM. On the other hand, iron is continuously delivered by SNe~Ia resulting in a decrement of the [$\alpha$/Fe] ratios through younger ages.

Like iron, the iron-peak elements V, Mn, Co, and Ni are produced by both SNe~II and SNe~Ia, and their [X/Fe] vs age relations are similar to that displayed by Na: an early [X/Fe] rise, the knee at $\sim$5~Gyr, and a later decrement towards younger ages (see Fig.~\ref{XFe_age}). For these elements, the [X/Fe] increment extended in time likely reflects the longer timescales in which SNe~Ia pollute the ISM. However, the reason for the knee and the late [X/Fe] decrease are still unclear. The significant, and in some cases dominant, contribution from SNe~Ia toward the production of these elements excludes a simple explanation, mainly because of the variety of the possible SN~Ia sources. 

A series of observational studies made clear that both the environment and the path that lead to SNe~Ia explosions play a role in their yields (e.g., \citealt{Stritzinger06a,Ellis08,Howell09,Kelly10,Foley13,Graham15}). These results corroborate the outcomes of several theoretical models suggesting that the SNe~Ia yields are strongly affected by the metallicity, mass, and age of the progenitor (e.g., \citealt{Timmes03,Travaglio05,Mannucci06,Jackson10,De14,Travaglio15}), as well as by asymmetries in the explosion \citep{Chamulak12}. 

It is plausible that the late evolution of the iron-peak elements could be the result of the different two main channels that trigger the SN~Ia explosions: the single- and the double-degenerate scenarios \citep{Iben84,Webbink84}. A key difference, for example, is the fact that the two channels involve very different time distributions of the SNe rates (e.g., \citealt{Yungelson00,Ruiter09,Mennekens10,Yungelson10,Bours13,Wang13,Claeys14,Maoz14}). Namely, for the single-degenerate model, most of the possible donors$\footnote{The single-degenerate scenario for near-Chandrasekhar mass white dwarfs probably has at least two sub-channels dependent on the mass of the donor \citep{Kobayashi98}. A first class involves helium-burning intermediate-mass stars donating mass to the white dwarf, while in a second class the donor is a red giant or main sequence star of $\sim$1M$_{\sun}$. The first class has shorter timescales ($\sim$100-500~Myr) and it is the dominant channel, while the second has timescales $>$1~Gyr \citep{Ruiter14}.}$ are limited in a strict range of masses, $\sim$2-3~M$_{\sun}$ \citep{Langer00,Han04}, and since these systems explode shortly after the donor evolves of the main sequence, the resulting SNe~Ia explosions are mostly concentrated in a narrow interval of time between a few hundred of Myrs and 2 Gyr after an onset of star formation. 
On the other hand, besides the prompt SNe~Ia explosions proposed by \citet{Bonaparte13}, the merger rate in the double-degenerate models is expected to reach a maximum very early (the typical time in which the stars with masses below 8~M$_{\sun}$ evolve into the first white dwarfs) and then moderately declines with time. This implies that the production of iron-elements in the early evolution of the Galaxy should be driven by both the SN~Ia channels, but after $\sim$5~Gyr from the initial burst of star formation it should be dominated by SNe~Ia caused by the merger of white dwarfs.


It is also very likely that the exploding white dwarf mass in the double-degenerate model, might play an important role in the chemical evolution of the Galactic disk. 
In addition, it is generally accepted that, under certain circumstances, the SN~Ia detonation can be activated even if the combined mass does not exceed the Chandrasekhar mass (e.g., \citealt{Pakmor10,vanKerkwijk10}). Owing to the lower mass of the progenitors, this $sub$-Chandrasekhar mass scenario could be of great importance in the latest evolution of the Galaxy. It is also remarkable that, as shown by \citet{Seitenzahl13a}, the material released into the ISM by the $sub$-Chandrasekhar SN~Ia have a [Mn/Fe] ratio that is negative and significantly lower than that produced by $near$-Chandrasekhar SN~Ia explosions. This would imply a late decrement of the [Mn/Fe] ratio as time goes on, similar to what we observe in Fig.~\ref{XFe_age}.

Finally, the recent calculations performed by \citet{Miles15} have shown that the variations in the progenitor metallicity results in several trends in the SNe~Ia yields, since they are affected by the neutron excess in the composition of the progenitor white dwarf. Namely, the increase of the initial metallicity would lead to a more efficient production of iron (see Fig. 3 in \citealt{Miles15}). This could account for the lower [X/Fe] ratios observed for the iron-peak elements in the youngest stars. 

\begin{figure*}
\centering
\includegraphics[width=17cm]{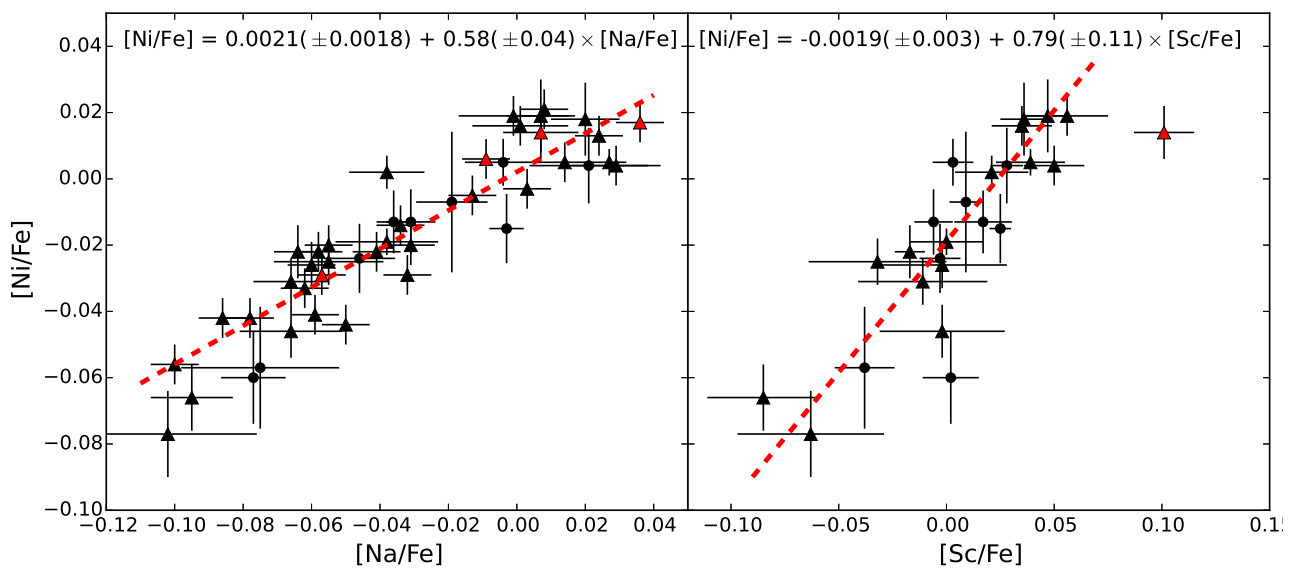} \caption{[Ni/Fe] as a function of [Na/Fe] (left) and as a function of [Sc/Fe] (right), with the symbols as in Fig.~\ref{XFe_age}. The dashed red lines correspond to linear fits.}
\label{Ni_Na}
\end{figure*}

Additional hints on the production of an iron-peak element like Ni could be provided by the plot in Fig.~\ref{Ni_Na} (left) showing the existence of a tight relation between the [Na/Fe] and [Ni/Fe] ratios. A linear fit that takes into account the errors in both coordinates provides the following relation:
$$\left[ \frac{Ni}{Fe} \right]=0.0021\left(\pm0.0018\right)+0.58\left(\pm0.04\right)\times\left[ \frac{Na}{Fe} \right],$$
with a standard deviation of $\sigma$$_{[Ni/Fe]}$=0.008~dex. This correlation was already seen among solar twin stars (N15) and for stars in dwarf spheroidal galaxies \citep{Venn04,Letarte10,Lemasle14}. As enlightened by \citet{Venn04}, the $^{23}$Na is the only stable neutron-rich isotope produced in large amounts by the hydrostatic C burning during the SN~II events. Also the stable $^{58}$Ni isotope is yielded during the SN~II explosions, but its production is controlled by the abundance of the neutron-rich elements, like $^{23}$Na; for more details see \citet{Thielemann90}. This is why a Na-Ni relation is somehow expected when the chemical enrichment is dominated by SN~II explosions. However, Ni is significantly produced also by SN~Ia explosions, whose Na yields are not well constrained. This is why it is still unclear if there is a contribution of SNe~Ia to the Na-Ni relation. Anyway, our data clearly show that such relation exist for stars of all the ages and of the two populations (thin and thick disk), which may suggests that the channels that are leading the latest chemical evolution of the Galactic disk are maintaining the Na-Ni dependence.

Other elements, such as Sc, show similar correlations with Ni (Fig.~\ref{Ni_Na} - right) due to the neutron-rich environment that boosts the production of both the elements (e.g., \citealt{Kobayashi06}). However, the Sc-Ni correlation is not as tight as that found for Na-Ni. In addition, contrarily to the Na-Ni relation, the thick-disk star is an outlier of the Sc-Ni main distribution and, for this reason, has not been considered for the linear fit.

Also the mechanism that is driving the increase of the [Ca/Fe] and [Cr/Fe] ratios seen for the youngest stars, largely remains an open question. Probably, as for the iron-peak elements, the models' uncertainties and the large variety of the SNe~Ia explosion mechanisms, complicate a simple explanation. For instance, as suggested by N15, the increasing trend as time goes on observed for Ca could be the effect of a new type of SNe characterised by a large production of Ca that would be up to 10 times greater than that synthesised by SN~Ia \citep{Perets10}.

\begin{figure*}
\centering
\includegraphics[width=17cm]{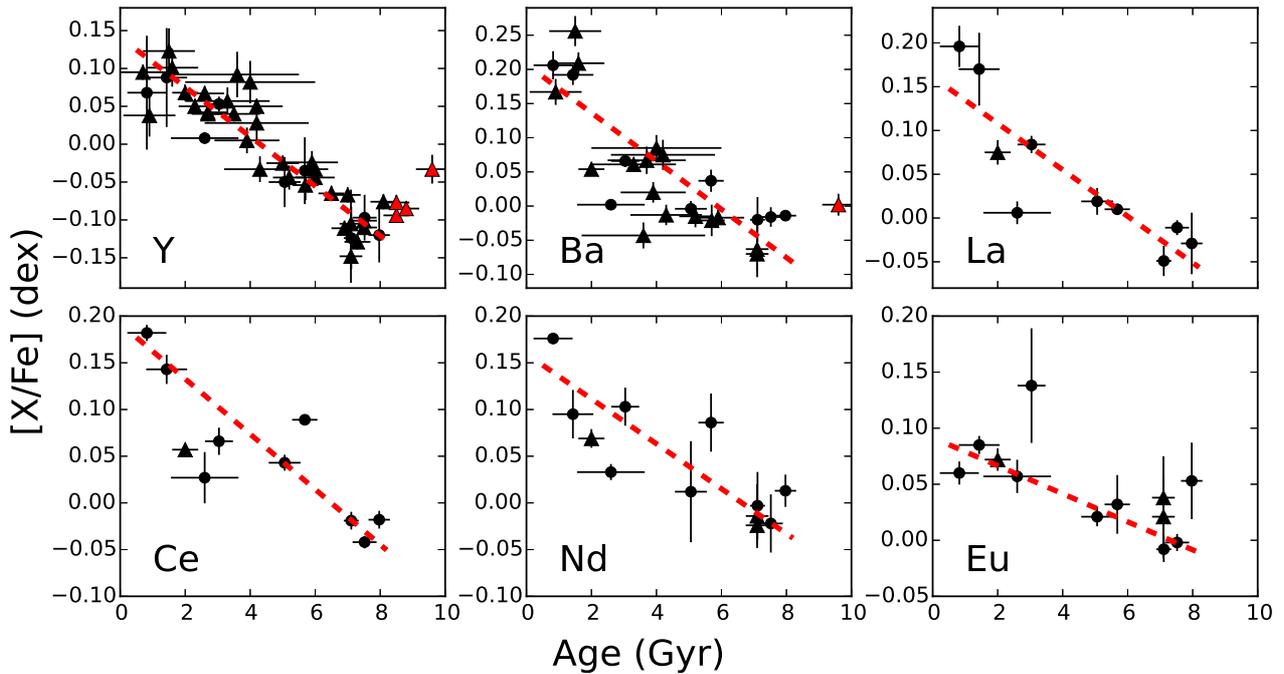} \caption{[X/Fe] ratios as a function of stellar ages for the nine solar twins analysed here (circles). We also plotted the measurements from \citet{Ramirez11,Nissen15,Spina16,YanaGalarza16a}, as triangles. The red triangles represent the thick disk stars. The [X/Fe]-age distributions have been fitted by the functions showed as red dashed lines, whose parameters are listed in Table~\ref{age_fit_he}.}
\label{XFe_age_he}
\end{figure*}

\subsection{The neutron-capture elements}

The [X/Fe] ratios of the $n$-capture elements are plotted in Fig.~\ref{XFe_age_he} as a function of the stellar ages. As mentioned in Section~\ref{Intr}, these species are synthesised by a mixture of the $s$- and $r$-process. The thin disk stars are characterised by a [X/Fe] increment as time goes on, which is probably the result of the elemental synthesis from the $s$-process in AGB stars. In fact, the latest contribution to chemical evolution by AGB stars, especially those of lower mass \citep{Busso01,DOrazi09c,Maiorca11}, should have an impact on the production $n$-capture elements greater than that of SNe \citep{Battistini16}. Therefore, the [X/Fe]-age dependency of each $n$-element is probably mainly controlled in the thin disk by its rate of production through the $s$-process, rather than the $r$-process.

\begin{figure}
\centering
\includegraphics[width=9cm]{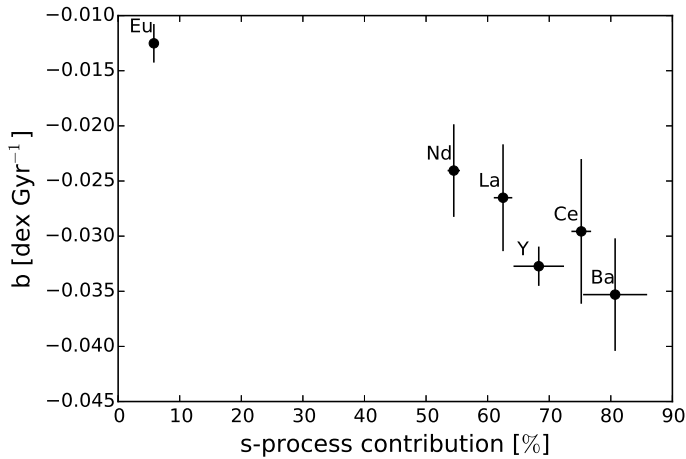} \caption{Slopes of the [X/Fe]-age relations found for the $n$-capture elements (i.e., the b parameters in Table~\ref{age_fit_he}) as a function of their solar $s$-process contribution percentages taken from \citet{Bisterzo14}.}
\label{sprocess}
\end{figure}

With the aim to test this hypothesis, we performed a linear fit of the [X/Fe] vs age relations shown by the thin disk stars in Fig.~\ref{XFe_age_he}, adopting the orthogonal distance regression method and the following formula, [X/Fe]$=$a+b$\times$age. The resulting parameters are listed in Table~\ref{age_fit_he} and the linear fits are over-plotted in Fig.~\ref{XFe_age_he}. In Fig.~\ref{sprocess} we show the slope of each element (i.e., the b parameters), obtained by the linear fitting, as a function of their solar $s$-process contribution percentages taken from \citet{Bisterzo14}. 
Based on these production rates, we can consider Ba as a ``prototype'' $s$-process element owing to its high $s$-process contribution, which is responsible for $\sim$81\% of Ba. On the other hand, Eu can be considered a nearly pure $r$-process element, since the $s$-process accounts for only $\sim$6$\%$ of its synthesis. The difference in the production rates through $s$-process between these two elements results in two very different b slopes: Ba has a very steep [X/Fe]-age dependence, while the correlation found for Eu is almost flat. Indeed, a relation between the b parameters and the $s$-process contributions exists in Fig.~\ref{sprocess} for all the $n$-capture elements. Namely, the elements with a higher contribution from the $s$-process are those with a steeper [X/Fe]-age slope. This result is consistent with AGBs being mainly responsible for the enrichment of $n$-capture elements in the thin disk. However, this hypothesis assumes that all the $r$-process production occurs on sites of nucleosynthesis of very short timescales (e.g., SNe~II) and that the contribution of the $r$-process is negligible in the latest stage of galactic evolution. This hypothesis is not necessarily true since colliding neutron stars can produce $r$-process elements (e.g., \citealt{Wanajo14}), although notice that the timescales of these mergers are also short since a larger fraction ($\sim$70$\%$) have a coalescence timescale $<$1~Gyr \citep{Vangioni16}.

Since the AGB yields are highly dependent on the progenitor mass \citep{Fishlock14} and on metallicity \citep{Busso01,Karakas16}, one would expect to observe non-linear [X/Fe]-age relations for the $s$-process elements resulting from the delayed pollution from low-mass AGB stars with respect to the early contribution from the intermediate-mass AGB stars. 
Unfortunately the few abundance determinations for the $n$-capture elements in solar twins prevents us from estabilishing eventual non-linearities in the [X/Fe] vs age plots shown in Fig.~\ref{XFe_age_he}. Better constraints in these relations would ensure powerful diagnostics of the AGB progenitor mass distribution.

\begin{table}
\small\begin{center}
\caption{\label{age_fit_he} Results of the fitting of $n$-capture elements}
\begin{tabular}{c|cc|cc} 
\hline\hline 
Element & a & b & $\sigma_{[X/Fe]}$ & $\chi^{2}_{red}$ \\
 & [dex] & [dex Gyr$^{-1}$] & [dex] & \\ \hline
Y & 0.141 $\pm$ 0.010 & $-$0.033 $\pm$ 0.002 & 0.024 & 1.02\\
Ba & 0.207 $\pm$ 0.027 & $-$0.035 $\pm$ 0.005 & 0.044 & 4.22 \\
La & 0.161 $\pm$ 0.026 & $-$0.027 $\pm$ 0.005 & 0.030 & 2.87 \\
Ce & 0.192 $\pm$ 0.037 & $-$0.030 $\pm$ 0.007 & 0.032 & 9.15 \\
Nd & 0.160 $\pm$ 0.020 & $-$0.024 $\pm$ 0.004 & 0.029 & 3.22 \\
Eu & 0.092 $\pm$ 0.009 & $-$0.013 $\pm$ 0.002 & 0.022 & 1.25 \\

\hline\hline 
\end{tabular}
\end{center}
\end{table}

The thick disk stars have [Y/Fe] ratios that are slightly greater than the ratios determined for most of the oldest thin disk stars. The change in slopes for the two different populations was already noted by \citet{Battistini16} for a number of $n$-capture elements. They argue that the [X/Fe] increase that they observed for Sr, Zr, La, Nd, Sm, and Eu in stars older than 8~Gyr, is due to the production of these elements via SNe~II, but it could also be due to massive stars (e.g., \citealt{Hansen14,Frischknecht16}).  

 \subsection{Chemical clocks}

\begin{figure*}
\centering
\includegraphics[width=17cm]{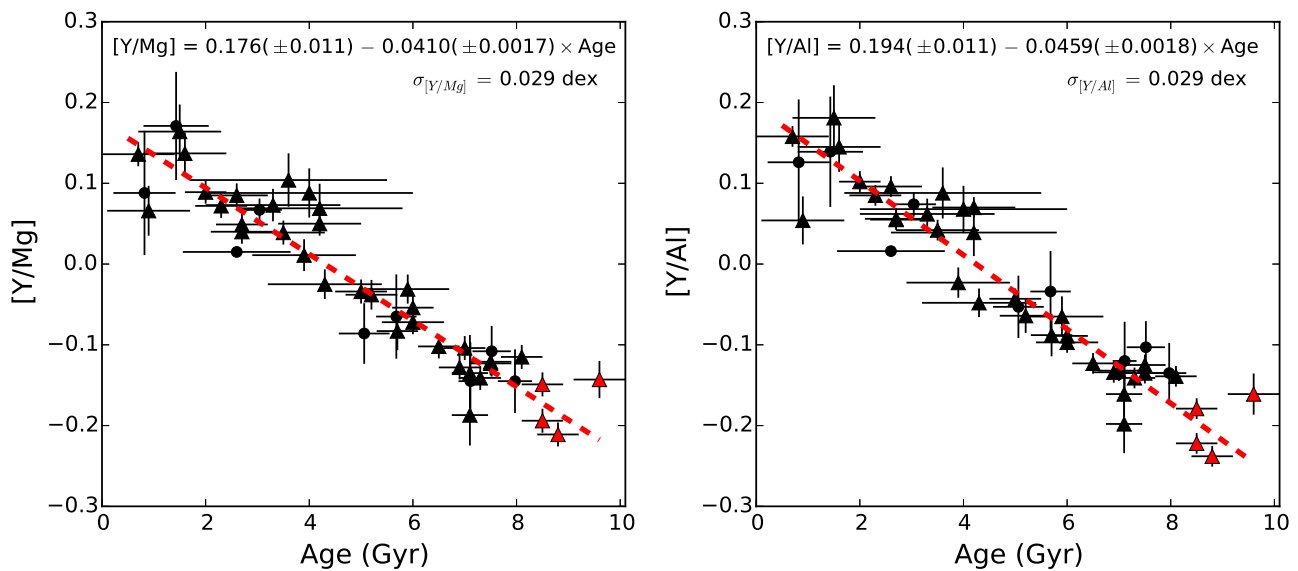} \caption{[Y/Mg] and [Y/Al] as a function of age, with the same symbols of Fig.~\ref{XFe_age}. The red dashed lines correspond to linear fits.}
\label{age_indicators}
\end{figure*}

The plot shown in Fig.~\ref{age_indicators} confirms that, as already found by N15 and \citet{TucciMaia16}, the [Y/Mg] ratio is a sensitive indicator of age. A linear fit that includes both the thin and thick disk stars and which takes into account the errors in both the coordinates, results in the following relation:
$$\left[ \frac{Y}{Mg} \right]=0.176\left(\pm0.011\right)-0.0410\left(\pm0.0017\right)\times Age\left[Gyr \right],$$
which is in excellent agreement with the relations provided by N15 and \citet{TucciMaia16}. However, it is noteworthy that also [Y/Al] shows a very tight correlation with age and that, most importantly, this [Y/Al] - age dependence is steeper than that exhibited by [Y/Mg]:
$$\left[ \frac{Y}{Al} \right]=0.194\left(\pm0.011\right)-0.0459\left(\pm0.0018\right)\times Age\left[Gyr \right].$$

The two relations have similar level of scatter (i.e., $\sigma$$_{[Y/Mg]}$=0.029 and $\sigma$$_{[Y/Al]}$=0.029~dex), thus both the [Y/Mg] and [Y/Al] ratios can serve as sensitive age indicators that are completely independent from the classical methods used for age determinations (e.g., isochrones, Li depletion, gyrochronology). Notice that the slightly higher steepness of the [Y/Al] - age relation would provide better constraints on the stellar ages with respect to the [Y/Mg] clock.

%

\section{Summary and conclusions}
\label{conclusions}
Using high-quality UVES spectra of nine solar twins, we obtained precise estimates of stellar ages and chemical abundances. These determinations have been integrated with additional accurate age and abundance determinations from recent spectroscopy studies of 32 solar twins existing in the literature \citep{Ramirez11,TucciMaia14,Nissen15,Spina16,YanaGalarza16a}. So far, this represent the largest data set from which high-precision abundances of 24 species have been determined in order to outline the [X/Fe]-age relations over a time interval of $\sim$10 Gyr.

Our main results can be summarised as follows:
\begin{itemize}
\item the thin and thick disk populations are clearly distinguishable by both the stellar ages and chemical composition (see Fig.~\ref{XFe_age} and \ref{age_fit_he}). All the thick disk stars are older than 8~Gyr and are enhanced in the $\alpha$-elements with respect to the thin disk population;
\item among the thin disk stars, tight correlations exist between the [X/Fe] ratios and the stellar ages for many elements that we analysed (see Fig.~\ref{XFe_age}). 
We fitted the [X/Fe]-age relation of the light elements (i.e., from C to Zn) using three different models based on a straight line, hyperbolic and two-segmented line functions, respectively. Through a statistical comparison of the results, we demonstrated that most of these species present non-linear or non-monotonic [X/Fe]-age relations. We showed that the hyperbolic model is the most appropriate in fitting the [X/Fe]-ages distribution of most of the light elements. This hyperbolic fit provides age estimates of the knees of the [X/Fe]-age relations. Nevertheless, the [X/Fe] ratios of C, O, Al, and Ti show a more linear dependence to age;
\item a relation connects the age at the knees of each species and their logarithmic ratios between the SN~II and SN~Ia yields (see Fig.~\ref{yields}). This relationship reflects the different histories of nucleosynthesis that each element underwent in the thin disk;
\item the stars with ages $\sim$6-8~Gyr are the most enriched in $\alpha$-elements among the thin disk population (Fig.~\ref{XFe_age}). This early enrichment in the species mostly produced by SNe~II, could be due to an initial and intensive star formation burst that gave origin to the thin disk. For stars younger than 6-7~Gyr, the [X/Fe] ratios of the $\alpha$-elements linearly declines with time. This decrement reflects the contribution of SNe~Ia to the chemical evolution of the disk, that started to be non-negligible $\sim$1~Gyr after the initial stellar formation burst;
\item like the $\alpha$-elements, Na is mainly produced by SNe~II. However it shows an early [X/Fe] increase that is somewhat more protracted in time with respect to the $\alpha$-elements (Fig.~\ref{XFe_age}). This peculiarity could be due to the high metallicity dependency of the Na yields from SNe~II;
\item the species that are synthesised by SNe~Ia present an early [X/Fe] increment that is extended in time (see Fig.~\ref{XFe_age}) and that could be the effect of the $\sim$1~Gyr delay in the contribution from SNe~Ia from the star formation burst. This early [X/Fe] increase ends at 4-6~Gyr and it is followed by a [X/Fe] decrease towards younger stars. The evolution of the iron-peak elements may be interpreted invoking multiple types of SNe~Ia, which are triggered by different progenitors and operate with different timescales;
\item the elements Ca and Cr are characterised by having a [X/Fe] increase for the youngest stars as time goes on (Fig.~\ref{XFe_age}). These trends are also difficult to interpret based on the classical yields of SNe~II and Ia. For instance, it is striking that Ca and the $\alpha$-elements show opposite behaviours;
\item the [X/Fe] ratios of the $n$-capture elements increase with time towards the youngest stars (see Fig.~\ref{age_fit_he}). The slopes of the [X/Fe]-age relations are related to the contribution percentages to each element from the $s$-process (see Fig~\ref{sprocess}): the elements that are mainly produced by the $s$-process are those with a steeper slope. This suggests that the enrichment in $n$-capture elements of the thin disk is mainly operated by AGB stars;
\item both the [Y/Mg] and [Y/Al] ratios are powerful age indicators (see Fig.~\ref{age_indicators}). However, among these two, the [Y/Al] ratio may be the best stellar clock since its dependence with age is slightly steeper than that found for [Y/Mg].
\end{itemize}

Knowledge of the [X/Fe]-age relations is a gold mine from which we can achieve a great understanding about the processes that governed the formation and evolution of the Milky Way: the nature of the star formation history, the SNe rates, the stellar yields, and the variety of the SNe progenitors, etc. Studies like that of N15 and this work pave the way to a reverse engineering effort, that is both timely and necessary to reconstruct, through chemical evolution models (e.g., \citealt{Chiappini97,Matteucci09,Romano10,Kobayashi11,Minchev14,Andrews16}), the nucleosynthetic history of the elements in the Galactic disk.

\begin{acknowledgements}
The authors are grateful to the anonymous referee, who made several important suggestions which have resulted in an improved body of work. We also thank Ashley Ruiter for comments on Type~Ia supernovae. LS and JM acknowledge the support from FAPESP (2014/15706-9 and 2012/24392-2). MA and DY acknowledge support from the Australian Research Council (grants FL110100012, DP120100991 and  FT140100554).
\end{acknowledgements}

%
%

\bibliographystyle{aa}
\bibliography{/Users/lspina/Dropbox/papers/bibliography.bib}

\begin{thebibliography}{123}
\expandafter\ifx\csname natexlab\endcsname\relax\def\natexlab#1{#1}\fi

\bibitem[{{Adibekyan} {et~al.}(2011){Adibekyan}, {Santos}, {Sousa}, \&
  {Israelian}}]{Adibekyan11}
{Adibekyan}, V.~Z., {Santos}, N.~C., {Sousa}, S.~G., \& {Israelian}, G. 2011,
  \aap, 535, L11

\bibitem[{{Amarsi} {et~al.}(2016){Amarsi}, {Asplund}, {Collet}, \&
  {Leenaarts}}]{Amarsi16}
{Amarsi}, A.~M., {Asplund}, M., {Collet}, R., \& {Leenaarts}, J. 2016, \mnras,
  455, 3735

\bibitem[{{Andrews} {et~al.}(2016){Andrews}, {Weinberg}, {Sch{\"o}nrich}, \&
  {Johnson}}]{Andrews16}
{Andrews}, B.~H., {Weinberg}, D.~H., {Sch{\"o}nrich}, R., \& {Johnson}, J.~A.
  2016, ArXiv e-prints [\eprint[arXiv]{1604.08613}]

\bibitem[{{Argast} {et~al.}(2004){Argast}, {Samland}, {Thielemann}, \&
  {Qian}}]{Argast04}
{Argast}, D., {Samland}, M., {Thielemann}, F.-K., \& {Qian}, Y.-Z. 2004, \aap,
  416, 997

\bibitem[{{Arlandini} {et~al.}(1999){Arlandini}, {K{\"a}ppeler}, {Wisshak},
  {Gallino}, {Lugaro}, {Busso}, \& {Straniero}}]{Arlandini99}
{Arlandini}, C., {K{\"a}ppeler}, F., {Wisshak}, K., {et~al.} 1999, \apj, 525,
  886

\bibitem[{{Battistini} \& {Bensby}(2016)}]{Battistini16}
{Battistini}, C. \& {Bensby}, T. 2016, \aap, 586, A49

\bibitem[{{Bedell} {et~al.}(2014){Bedell}, {Mel{\'e}ndez}, {Bean},
  {Ram{\'{\i}}rez}, {Leite}, \& {Asplund}}]{Bedell14}
{Bedell}, M., {Mel{\'e}ndez}, J., {Bean}, J.~L., {et~al.} 2014, \apj, 795, 23

\bibitem[{{Bensby} {et~al.}(2014){Bensby}, {Feltzing}, \& {Oey}}]{Bensby14}
{Bensby}, T., {Feltzing}, S., \& {Oey}, M.~S. 2014, \aap, 562, A71

\bibitem[{{Bensby} {et~al.}(2007){Bensby}, {Zenn}, {Oey}, \&
  {Feltzing}}]{Bensby07}
{Bensby}, T., {Zenn}, A.~R., {Oey}, M.~S., \& {Feltzing}, S. 2007, \apjl, 663,
  L13

\bibitem[{{Bergemann} {et~al.}(2012){Bergemann}, {Lind}, {Collet}, {Magic}, \&
  {Asplund}}]{Bergemann12}
{Bergemann}, M., {Lind}, K., {Collet}, R., {Magic}, Z., \& {Asplund}, M. 2012,
  \mnras, 427, 27

\bibitem[{{Biazzo} {et~al.}(2015){Biazzo}, {Gratton}, {Desidera}, {Lucatello},
  {Sozzetti}, {Bonomo}, {Damasso}, {Gandolfi}, {Affer}, {Boccato}, {Borsa},
  {Claudi}, {Cosentino}, {Covino}, {Knapic}, {Lanza}, {Maldonado}, {Marzari},
  {Micela}, {Molaro}, {Pagano}, {Pedani}, {Pillitteri}, {Piotto}, {Poretti},
  {Rainer}, {Santos}, {Scandariato}, \& {Zanmar Sanchez}}]{Biazzo15}
{Biazzo}, K., {Gratton}, R., {Desidera}, S., {et~al.} 2015, ArXiv e-prints
  [\eprint[arXiv]{1506.01614}]

\bibitem[{{Bisterzo} {et~al.}(2014){Bisterzo}, {Travaglio}, {Gallino},
  {Wiescher}, \& {K{\"a}ppeler}}]{Bisterzo14}
{Bisterzo}, S., {Travaglio}, C., {Gallino}, R., {Wiescher}, M., \&
  {K{\"a}ppeler}, F. 2014, \apj, 787, 10

\bibitem[{Boggs \& Rogers(1990)}]{Boggs90}
Boggs, P.~T. \& Rogers, J.~E. 1990, Contemporary Mathematics, 112, 183

\bibitem[{{Bonaparte} {et~al.}(2013){Bonaparte}, {Matteucci}, {Recchi},
  {Spitoni}, {Pipino}, \& {Grieco}}]{Bonaparte13}
{Bonaparte}, I., {Matteucci}, F., {Recchi}, S., {et~al.} 2013, \mnras, 435,
  2460

\bibitem[{{Bours} {et~al.}(2013){Bours}, {Toonen}, \& {Nelemans}}]{Bours13}
{Bours}, M.~C.~P., {Toonen}, S., \& {Nelemans}, G. 2013, \aap, 552, A24

\bibitem[{{Busso} {et~al.}(1999){Busso}, {Gallino}, \& {Wasserburg}}]{Busso99}
{Busso}, M., {Gallino}, R., \& {Wasserburg}, G.~J. 1999, \araa, 37, 239

\bibitem[{{Busso} {et~al.}(2001){Busso}, {Marengo}, {Travaglio}, {Corcione}, \&
  {Silvestro}}]{Busso01}
{Busso}, M., {Marengo}, M., {Travaglio}, C., {Corcione}, L., \& {Silvestro}, G.
  2001, \memsai, 72, 309

\bibitem[{{Castelli} \& {Kurucz}(2004)}]{Castelli04}
{Castelli}, F. \& {Kurucz}, R.~L. 2004, ArXiv Astrophysics e-prints
  [\eprint{astro-ph/0405087}]

\bibitem[{{Cayrel} \& {Cayrel de Strobel}(1966)}]{Cayrel66}
{Cayrel}, R. \& {Cayrel de Strobel}, G. 1966, \araa, 4, 1

\bibitem[{{Chamulak} {et~al.}(2012){Chamulak}, {Meakin}, {Seitenzahl}, \&
  {Truran}}]{Chamulak12}
{Chamulak}, D.~A., {Meakin}, C.~A., {Seitenzahl}, I.~R., \& {Truran}, J.~W.
  2012, \apj, 744, 27

\bibitem[{{Chiappini} {et~al.}(1997){Chiappini}, {Matteucci}, \&
  {Gratton}}]{Chiappini97}
{Chiappini}, C., {Matteucci}, F., \& {Gratton}, R. 1997, \apj, 477, 765

\bibitem[{{Claeys} {et~al.}(2014){Claeys}, {Pols}, {Izzard}, {Vink}, \&
  {Verbunt}}]{Claeys14}
{Claeys}, J.~S.~W., {Pols}, O.~R., {Izzard}, R.~G., {Vink}, J., \& {Verbunt},
  F.~W.~M. 2014, \aap, 563, A83

\bibitem[{{Cowan} \& {Sneden}(2004)}]{Cowan04}
{Cowan}, J.~J. \& {Sneden}, C. 2004, Origin and Evolution of the Elements, 27

\bibitem[{{De} {et~al.}(2014){De}, {Timmes}, {Brown}, {Calder}, {Townsley},
  {Athanassiadou}, {Chamulak}, {Hawley}, \& {Jack}}]{De14}
{De}, S., {Timmes}, F.~X., {Brown}, E.~F., {et~al.} 2014, \apj, 787, 149

\bibitem[{{D'Orazi} {et~al.}(2009){D'Orazi}, {Magrini}, {Randich}, {Galli},
  {Busso}, \& {Sestito}}]{DOrazi09c}
{D'Orazi}, V., {Magrini}, L., {Randich}, S., {et~al.} 2009, \apjl, 693, L31

\bibitem[{{Edvardsson} {et~al.}(1993){Edvardsson}, {Andersen}, {Gustafsson},
  {Lambert}, {Nissen}, \& {Tomkin}}]{Edvardsson93}
{Edvardsson}, B., {Andersen}, J., {Gustafsson}, B., {et~al.} 1993, \aap, 275,
  101

\bibitem[{{Ellis} {et~al.}(2008){Ellis}, {Sullivan}, {Nugent}, {Howell},
  {Gal-Yam}, {Astier}, {Balam}, {Balland}, {Basa}, {Carlberg}, {Conley},
  {Fouchez}, {Guy}, {Hardin}, {Hook}, {Pain}, {Perrett}, {Pritchet}, \&
  {Regnault}}]{Ellis08}
{Ellis}, R.~S., {Sullivan}, M., {Nugent}, P.~E., {et~al.} 2008, \apj, 674, 51

\bibitem[{{Epstein} {et~al.}(2010){Epstein}, {Johnson}, {Dong}, {Udalski},
  {Gould}, \& {Becker}}]{Epstein10}
{Epstein}, C.~R., {Johnson}, J.~A., {Dong}, S., {et~al.} 2010, \apj, 709, 447

\bibitem[{{Fabbian} {et~al.}(2009){Fabbian}, {Asplund}, {Barklem}, {Carlsson},
  \& {Kiselman}}]{Fabbian09}
{Fabbian}, D., {Asplund}, M., {Barklem}, P.~S., {Carlsson}, M., \& {Kiselman},
  D. 2009, \aap, 500, 1221

\bibitem[{{Fisher}(1925)}]{Fisher25}
{Fisher}, R. 1925, Oliver and Boyd, Edinburgh, UK

\bibitem[{{Fishlock} {et~al.}(2014){Fishlock}, {Karakas}, {Lugaro}, \&
  {Yong}}]{Fishlock14}
{Fishlock}, C.~K., {Karakas}, A.~I., {Lugaro}, M., \& {Yong}, D. 2014, \apj,
  797, 44

\bibitem[{{Foley} {et~al.}(2013){Foley}, {Challis}, {Chornock},
  {Ganeshalingam}, {Li}, {Marion}, {Morrell}, {Pignata}, {Stritzinger},
  {Silverman}, {Wang}, {Anderson}, {Filippenko}, {Freedman}, {Hamuy}, {Jha},
  {Kirshner}, {McCully}, {Persson}, {Phillips}, {Reichart}, \&
  {Soderberg}}]{Foley13}
{Foley}, R.~J., {Challis}, P.~J., {Chornock}, R., {et~al.} 2013, \apj, 767, 57

\bibitem[{{Frischknecht} {et~al.}(2016){Frischknecht}, {Hirschi}, {Pignatari},
  {Maeder}, {Meynet}, {Chiappini}, {Thielemann}, {Rauscher}, {Georgy}, \&
  {Ekstr{\"o}m}}]{Frischknecht16}
{Frischknecht}, U., {Hirschi}, R., {Pignatari}, M., {et~al.} 2016, \mnras, 456,
  1803

\bibitem[{{Gibson} {et~al.}(2003){Gibson}, {Fenner}, {Renda}, {Kawata}, \&
  {Lee}}]{Gibson03}
{Gibson}, B.~K., {Fenner}, Y., {Renda}, A., {Kawata}, D., \& {Lee}, H.-c. 2003,
  \pasa, 20, 401

\bibitem[{{Gilmore} {et~al.}(1989){Gilmore}, {Wyse}, \& {Kuijken}}]{Gilmore89}
{Gilmore}, G., {Wyse}, R.~F.~G., \& {Kuijken}, K. 1989, \araa, 27, 555

\bibitem[{{Graham} {et~al.}(2015){Graham}, {Nugent}, {Sullivan}, {Filippenko},
  {Cenko}, {Silverman}, {Clubb}, \& {Zheng}}]{Graham15}
{Graham}, M.~L., {Nugent}, P.~E., {Sullivan}, M., {et~al.} 2015, \mnras, 454,
  1948

\bibitem[{{Han} \& {Podsiadlowski}(2004)}]{Han04}
{Han}, Z. \& {Podsiadlowski}, P. 2004, \mnras, 350, 1301

\bibitem[{{Hansen} {et~al.}(2014){Hansen}, {Montes}, \& {Arcones}}]{Hansen14}
{Hansen}, C.~J., {Montes}, F., \& {Arcones}, A. 2014, \apj, 797, 123

\bibitem[{{Haywood} {et~al.}(2013){Haywood}, {Di Matteo}, {Lehnert}, {Katz}, \&
  {G{\'o}mez}}]{Haywood13}
{Haywood}, M., {Di Matteo}, P., {Lehnert}, M.~D., {Katz}, D., \& {G{\'o}mez},
  A. 2013, \aap, 560, A109

\bibitem[{{Hillebrandt} {et~al.}(2013){Hillebrandt}, {Kromer}, {R{\"o}pke}, \&
  {Ruiter}}]{Hillebrandt13}
{Hillebrandt}, W., {Kromer}, M., {R{\"o}pke}, F.~K., \& {Ruiter}, A.~J. 2013,
  Frontiers of Physics, 8, 116

\bibitem[{{Howell} {et~al.}(2009){Howell}, {Sullivan}, {Brown}, {Conley}, {Le
  Borgne}, {Hsiao}, {Astier}, {Balam}, {Balland}, {Basa}, {Carlberg},
  {Fouchez}, {Guy}, {Hardin}, {Hook}, {Pain}, {Perrett}, {Pritchet},
  {Regnault}, {Baumont}, {LeDu}, {Lidman}, {Perlmutter}, {Suzuki}, {Walker}, \&
  {Wheeler}}]{Howell09}
{Howell}, D.~A., {Sullivan}, M., {Brown}, E.~F., {et~al.} 2009, \apj, 691, 661

\bibitem[{{Iben} \& {Tutukov}(1984)}]{Iben84}
{Iben}, Jr., I. \& {Tutukov}, A.~V. 1984, \apj, 284, 719

\bibitem[{{Jackson} {et~al.}(2010){Jackson}, {Calder}, {Townsley}, {Chamulak},
  {Brown}, \& {Timmes}}]{Jackson10}
{Jackson}, A.~P., {Calder}, A.~C., {Townsley}, D.~M., {et~al.} 2010, \apj, 720,
  99

\bibitem[{{Jim{\'e}nez} {et~al.}(2015){Jim{\'e}nez}, {Tissera}, \&
  {Matteucci}}]{Jimenez15}
{Jim{\'e}nez}, N., {Tissera}, P.~B., \& {Matteucci}, F. 2015, \apj, 810, 137

\bibitem[{{Karakas}(2010)}]{Karakas10}
{Karakas}, A.~I. 2010, \mnras, 403, 1413

\bibitem[{{Karakas} \& {Lattanzio}(2014)}]{Karakas14}
{Karakas}, A.~I. \& {Lattanzio}, J.~C. 2014, \pasa, 31, e030

\bibitem[{{Karakas} \& {Lugaro}(2016)}]{Karakas16}
{Karakas}, A.~I. \& {Lugaro}, M. 2016, \apj, submitted

\bibitem[{{Kelly} {et~al.}(2010){Kelly}, {Hicken}, {Burke}, {Mandel}, \&
  {Kirshner}}]{Kelly10}
{Kelly}, P.~L., {Hicken}, M., {Burke}, D.~L., {Mandel}, K.~S., \& {Kirshner},
  R.~P. 2010, \apj, 715, 743

\bibitem[{{Kim} {et~al.}(2002){Kim}, {Demarque}, {Yi}, \& {Alexander}}]{Kim02}
{Kim}, Y.-C., {Demarque}, P., {Yi}, S.~K., \& {Alexander}, D.~R. 2002, \apjs,
  143, 499

\bibitem[{{Kobayashi} \& {Nakasato}(2011)}]{Kobayashi11}
{Kobayashi}, C. \& {Nakasato}, N. 2011, \apj, 729, 16

\bibitem[{{Kobayashi} {et~al.}(1998){Kobayashi}, {Tsujimoto}, {Nomoto},
  {Hachisu}, \& {Kato}}]{Kobayashi98}
{Kobayashi}, C., {Tsujimoto}, T., {Nomoto}, K., {Hachisu}, I., \& {Kato}, M.
  1998, \apjl, 503, L155

\bibitem[{{Kobayashi} {et~al.}(2006){Kobayashi}, {Umeda}, {Nomoto}, {Tominaga},
  \& {Ohkubo}}]{Kobayashi06}
{Kobayashi}, C., {Umeda}, H., {Nomoto}, K., {Tominaga}, N., \& {Ohkubo}, T.
  2006, \apj, 653, 1145

\bibitem[{{Korobkin} {et~al.}(2012){Korobkin}, {Rosswog}, {Arcones}, \&
  {Winteler}}]{Korobkin12}
{Korobkin}, O., {Rosswog}, S., {Arcones}, A., \& {Winteler}, C. 2012, \mnras,
  426, 1940

\bibitem[{{Kratz} {et~al.}(2007){Kratz}, {Farouqi}, {Pfeiffer}, {Truran},
  {Sneden}, \& {Cowan}}]{Kratz07}
{Kratz}, K.-L., {Farouqi}, K., {Pfeiffer}, B., {et~al.} 2007, \apj, 662, 39

\bibitem[{{Langer} {et~al.}(2000){Langer}, {Deutschmann}, {Wellstein}, \&
  {H{\"o}flich}}]{Langer00}
{Langer}, N., {Deutschmann}, A., {Wellstein}, S., \& {H{\"o}flich}, P. 2000,
  \aap, 362, 1046

\bibitem[{{Lemasle} {et~al.}(2014){Lemasle}, {de Boer}, {Hill}, {Tolstoy},
  {Irwin}, {Jablonka}, {Venn}, {Battaglia}, {Starkenburg}, {Shetrone},
  {Letarte}, {Fran{\c c}ois}, {Helmi}, {Primas}, {Kaufer}, \&
  {Szeifert}}]{Lemasle14}
{Lemasle}, B., {de Boer}, T.~J.~L., {Hill}, V., {et~al.} 2014, \aap, 572, A88

\bibitem[{{Letarte} {et~al.}(2010){Letarte}, {Hill}, {Tolstoy}, {Jablonka},
  {Shetrone}, {Venn}, {Spite}, {Irwin}, {Battaglia}, {Helmi}, {Primas},
  {Fran{\c c}ois}, {Kaufer}, {Szeifert}, {Arimoto}, \& {Sadakane}}]{Letarte10}
{Letarte}, B., {Hill}, V., {Tolstoy}, E., {et~al.} 2010, \aap, 523, A17

\bibitem[{{Lind} {et~al.}(2011){Lind}, {Asplund}, {Barklem}, \&
  {Belyaev}}]{Lind11}
{Lind}, K., {Asplund}, M., {Barklem}, P.~S., \& {Belyaev}, A.~K. 2011, \aap,
  528, A103

\bibitem[{{Magrini} {et~al.}(2009){Magrini}, {Sestito}, {Randich}, \&
  {Galli}}]{Magrini09}
{Magrini}, L., {Sestito}, P., {Randich}, S., \& {Galli}, D. 2009, \aap, 494, 95

\bibitem[{{Maiorca} {et~al.}(2011){Maiorca}, {Randich}, {Busso}, {Magrini}, \&
  {Palmerini}}]{Maiorca11}
{Maiorca}, E., {Randich}, S., {Busso}, M., {Magrini}, L., \& {Palmerini}, S.
  2011, \apj, 736, 120

\bibitem[{{Mannucci} {et~al.}(2006){Mannucci}, {Della Valle}, \&
  {Panagia}}]{Mannucci06}
{Mannucci}, F., {Della Valle}, M., \& {Panagia}, N. 2006, \mnras, 370, 773

\bibitem[{{Maoz} {et~al.}(2014){Maoz}, {Mannucci}, \& {Nelemans}}]{Maoz14}
{Maoz}, D., {Mannucci}, F., \& {Nelemans}, G. 2014, \araa, 52, 107

\bibitem[{{Matteucci}(2014)}]{Matteucci14}
{Matteucci}, F. 2014, The Origin of the Galaxy and Local Group, Saas-Fee
  Advanced Course, Volume 37.~ISBN 978-3-642-41719-1.~Springer-Verlag Berlin
  Heidelberg, 37, 145

\bibitem[{{Matteucci} \& {Greggio}(1986)}]{Matteucci86}
{Matteucci}, F. \& {Greggio}, L. 1986, \aap, 154, 279

\bibitem[{{Matteucci} \& {Recchi}(2001)}]{Matteucci01}
{Matteucci}, F. \& {Recchi}, S. 2001, \apj, 558, 351

\bibitem[{{Matteucci} {et~al.}(2009){Matteucci}, {Spitoni}, {Recchi}, \&
  {Valiante}}]{Matteucci09}
{Matteucci}, F., {Spitoni}, E., {Recchi}, S., \& {Valiante}, R. 2009, \aap,
  501, 531

\bibitem[{{Mel{\'e}ndez} {et~al.}(2012){Mel{\'e}ndez}, {Bergemann}, {Cohen},
  {Endl}, {Karakas}, {Ram{\'{\i}}rez}, {Cochran}, {Yong}, {MacQueen},
  {Kobayashi}, \& {Asplund}}]{Melendez12}
{Mel{\'e}ndez}, J., {Bergemann}, M., {Cohen}, J.~G., {et~al.} 2012, \aap, 543,
  A29

\bibitem[{{Mel{\'e}ndez} \& {et al.}(2016)}]{Melendez16}
{Mel{\'e}ndez}, J. \& {et al.} 2016, \aap, submitted

\bibitem[{{Mel{\'e}ndez} {et~al.}(2014){Mel{\'e}ndez}, {Ram{\'{\i}}rez},
  {Karakas}, {Yong}, {Monroe}, {Bedell}, {Bergemann}, {Asplund}, {Tucci Maia},
  {Bean}, {do Nascimento}, {Bazot}, {Alves-Brito}, {Freitas}, \&
  {Castro}}]{Melendez14}
{Mel{\'e}ndez}, J., {Ram{\'{\i}}rez}, I., {Karakas}, A.~I., {et~al.} 2014,
  \apj, 791, 14

\bibitem[{{Mennekens} {et~al.}(2010){Mennekens}, {Vanbeveren}, {De Greve}, \&
  {De Donder}}]{Mennekens10}
{Mennekens}, N., {Vanbeveren}, D., {De Greve}, J.~P., \& {De Donder}, E. 2010,
  \aap, 515, A89

\bibitem[{{Miles} {et~al.}(2015){Miles}, {van Rossum}, {Townsley}, {Timmes},
  {Jackson}, {Calder}, \& {Brown}}]{Miles15}
{Miles}, B.~J., {van Rossum}, D.~R., {Townsley}, D.~M., {et~al.} 2015, ArXiv
  e-prints [\eprint[arXiv]{1508.05961}]

\bibitem[{{Minchev} {et~al.}(2014){Minchev}, {Chiappini}, \&
  {Martig}}]{Minchev14}
{Minchev}, I., {Chiappini}, C., \& {Martig}, M. 2014, \aap, 572, A92

\bibitem[{{Monroe} {et~al.}(2013){Monroe}, {Mel{\'e}ndez}, {Ram{\'{\i}}rez},
  {Yong}, {Bergemann}, {Asplund}, {Bedell}, {Tucci Maia}, {Bean}, {Lind},
  {Alves-Brito}, {Casagrande}, {Castro}, {do Nascimento}, {Bazot}, \&
  {Freitas}}]{Monroe13}
{Monroe}, T.~R., {Mel{\'e}ndez}, J., {Ram{\'{\i}}rez}, I., {et~al.} 2013,
  \apjl, 774, L32

\bibitem[{{Nissen}(2015)}]{Nissen15}
{Nissen}, P.~E. 2015, \aap, 579, A52

\bibitem[{{Nomoto} {et~al.}(1997{\natexlab{a}}){Nomoto}, {Hashimoto},
  {Tsujimoto}, {Thielemann}, {Kishimoto}, {Kubo}, \& {Nakasato}}]{Nomoto97a}
{Nomoto}, K., {Hashimoto}, M., {Tsujimoto}, T., {et~al.} 1997{\natexlab{a}},
  Nuclear Physics A, 616, 79

\bibitem[{{Nomoto} {et~al.}(1997{\natexlab{b}}){Nomoto}, {Iwamoto}, {Nakasato},
  {Thielemann}, {Brachwitz}, {Tsujimoto}, {Kubo}, \& {Kishimoto}}]{Nomoto97b}
{Nomoto}, K., {Iwamoto}, K., {Nakasato}, N., {et~al.} 1997{\natexlab{b}},
  Nuclear Physics A, 621, 467

\bibitem[{{Pakmor} {et~al.}(2010){Pakmor}, {Kromer}, {R{\"o}pke}, {Sim},
  {Ruiter}, \& {Hillebrandt}}]{Pakmor10}
{Pakmor}, R., {Kromer}, M., {R{\"o}pke}, F.~K., {et~al.} 2010, \nat, 463, 61

\bibitem[{{Pearson}(1900)}]{Pearson00}
{Pearson}, K. 1900, Philosophical Magazine, 5, 157

\bibitem[{{Perets} {et~al.}(2010){Perets}, {Gal-Yam}, {Mazzali}, {Arnett},
  {Kagan}, {Filippenko}, {Li}, {Arcavi}, {Cenko}, {Fox}, {Leonard}, {Moon},
  {Sand}, {Soderberg}, {Anderson}, {James}, {Foley}, {Ganeshalingam}, {Ofek},
  {Bildsten}, {Nelemans}, {Shen}, {Weinberg}, {Metzger}, {Piro}, {Quataert},
  {Kiewe}, \& {Poznanski}}]{Perets10}
{Perets}, H.~B., {Gal-Yam}, A., {Mazzali}, P.~A., {et~al.} 2010, \nat, 465, 322

\bibitem[{{Pipino} \& {Matteucci}(2009)}]{Pipino09}
{Pipino}, A. \& {Matteucci}, F. 2009, in IAU Symposium, Vol. 258, The Ages of
  Stars, ed. E.~E. {Mamajek}, D.~R. {Soderblom}, \& R.~F.~G. {Wyse}, 39--50

\bibitem[{{Ram{\'{\i}}rez} {et~al.}(2007){Ram{\'{\i}}rez}, {Allende Prieto}, \&
  {Lambert}}]{Ramirez07}
{Ram{\'{\i}}rez}, I., {Allende Prieto}, C., \& {Lambert}, D.~L. 2007, \aap,
  465, 271

\bibitem[{{Ram{\'{\i}}rez} {et~al.}(2013){Ram{\'{\i}}rez}, {Allende Prieto}, \&
  {Lambert}}]{Ramirez13}
{Ram{\'{\i}}rez}, I., {Allende Prieto}, C., \& {Lambert}, D.~L. 2013, \apj,
  764, 78

\bibitem[{{Ram{\'{\i}}rez} {et~al.}(2012){Ram{\'{\i}}rez}, {Fish}, {Lambert},
  \& {Allende Prieto}}]{Ramirez12}
{Ram{\'{\i}}rez}, I., {Fish}, J.~R., {Lambert}, D.~L., \& {Allende Prieto}, C.
  2012, \apj, 756, 46

\bibitem[{{Ram{\'{\i}}rez} {et~al.}(2014{\natexlab{a}}){Ram{\'{\i}}rez},
  {Mel{\'e}ndez}, \& {Asplund}}]{Ramirez14}
{Ram{\'{\i}}rez}, I., {Mel{\'e}ndez}, J., \& {Asplund}, M. 2014{\natexlab{a}},
  \aap, 561, A7

\bibitem[{{Ram{\'{\i}}rez} {et~al.}(2014{\natexlab{b}}){Ram{\'{\i}}rez},
  {Mel{\'e}ndez}, {Bean}, {Asplund}, {Bedell}, {Monroe}, {Casagrande},
  {Schirbel}, {Dreizler}, {Teske}, {Tucci Maia}, {Alves-Brito}, \&
  {Baumann}}]{Ramirez14b}
{Ram{\'{\i}}rez}, I., {Mel{\'e}ndez}, J., {Bean}, J., {et~al.}
  2014{\natexlab{b}}, \aap, 572, A48

\bibitem[{{Ram{\'{\i}}rez} {et~al.}(2011){Ram{\'{\i}}rez}, {Mel{\'e}ndez},
  {Cornejo}, {Roederer}, \& {Fish}}]{Ramirez11}
{Ram{\'{\i}}rez}, I., {Mel{\'e}ndez}, J., {Cornejo}, D., {Roederer}, I.~U., \&
  {Fish}, J.~R. 2011, \apj, 740, 76

\bibitem[{{Recio-Blanco} {et~al.}(2014){Recio-Blanco}, {de Laverny},
  {Kordopatis}, {Helmi}, {Hill}, {Gilmore}, {Wyse}, {Adibekyan}, {Randich},
  {Asplund}, {Feltzing}, {Jeffries}, {Micela}, {Vallenari}, {Alfaro}, {Allende
  Prieto}, {Bensby}, {Bragaglia}, {Flaccomio}, {Koposov}, {Korn}, {Lanzafame},
  {Pancino}, {Smiljanic}, {Jackson}, {Lewis}, {Magrini}, {Morbidelli},
  {Prisinzano}, {Sacco}, {Worley}, {Hourihane}, {Bergemann}, {Costado},
  {Heiter}, {Joffre}, {Lardo}, {Lind}, \& {Maiorca}}]{RecioBlanco14}
{Recio-Blanco}, A., {de Laverny}, P., {Kordopatis}, G., {et~al.} 2014, \aap,
  567, A5

\bibitem[{{Reddy} {et~al.}(2006){Reddy}, {Lambert}, \& {Allende
  Prieto}}]{Reddy06}
{Reddy}, B.~E., {Lambert}, D.~L., \& {Allende Prieto}, C. 2006, \mnras, 367,
  1329

\bibitem[{{Romano} {et~al.}(2010){Romano}, {Karakas}, {Tosi}, \&
  {Matteucci}}]{Romano10}
{Romano}, D., {Karakas}, A.~I., {Tosi}, M., \& {Matteucci}, F. 2010, \aap, 522,
  A32

\bibitem[{{Ruiter} {et~al.}(2009){Ruiter}, {Belczynski}, \& {Fryer}}]{Ruiter09}
{Ruiter}, A.~J., {Belczynski}, K., \& {Fryer}, C. 2009, \apj, 699, 2026

\bibitem[{{Ruiter} {et~al.}(2014){Ruiter}, {Belczynski}, {Sim}, {Seitenzahl},
  \& {Kwiatkowski}}]{Ruiter14}
{Ruiter}, A.~J., {Belczynski}, K., {Sim}, S.~A., {Seitenzahl}, I.~R., \&
  {Kwiatkowski}, D. 2014, \mnras, 440, L101

\bibitem[{{Seitenzahl} {et~al.}(2013){Seitenzahl}, {Cescutti}, {R{\"o}pke},
  {Ruiter}, \& {Pakmor}}]{Seitenzahl13a}
{Seitenzahl}, I.~R., {Cescutti}, G., {R{\"o}pke}, F.~K., {Ruiter}, A.~J., \&
  {Pakmor}, R. 2013, \aap, 559, L5

\bibitem[{{Sneden}(1973)}]{Sneden73}
{Sneden}, C. 1973, \apj, 184, 839

\bibitem[{{Sneden} {et~al.}(2008){Sneden}, {Cowan}, \& {Gallino}}]{Sneden08}
{Sneden}, C., {Cowan}, J.~J., \& {Gallino}, R. 2008, \araa, 46, 241

\bibitem[{{Soubiran} {et~al.}(2013){Soubiran}, {Jasniewicz}, {Chemin}, {Crifo},
  {Udry}, {Hestroffer}, \& {Katz}}]{Soubiran13}
{Soubiran}, C., {Jasniewicz}, G., {Chemin}, L., {et~al.} 2013, \aap, 552, A64

\bibitem[{{Spina} {et~al.}(2016){Spina}, {Mel{\'e}ndez}, \&
  {Ram{\'{\i}}rez}}]{Spina16}
{Spina}, L., {Mel{\'e}ndez}, J., \& {Ram{\'{\i}}rez}, I. 2016, \aap, 585, A152

\bibitem[{{Stritzinger} {et~al.}(2006){Stritzinger}, {Leibundgut}, {Walch}, \&
  {Contardo}}]{Stritzinger06a}
{Stritzinger}, M., {Leibundgut}, B., {Walch}, S., \& {Contardo}, G. 2006, \aap,
  450, 241

\bibitem[{{Surman} {et~al.}(2008){Surman}, {McLaughlin}, {Ruffert}, {Janka}, \&
  {Hix}}]{Surman08}
{Surman}, R., {McLaughlin}, G.~C., {Ruffert}, M., {Janka}, H.-T., \& {Hix},
  W.~R. 2008, \apjl, 679, L117

\bibitem[{{Thielemann} {et~al.}(2011){Thielemann}, {Arcones}, {K{\"a}ppeli},
  {Liebend{\"o}rfer}, {Rauscher}, {Winteler}, {Fr{\"o}hlich}, {Dillmann},
  {Fischer}, {Martinez-Pinedo}, {Langanke}, {Farouqi}, {Kratz}, {Panov}, \&
  {Korneev}}]{Thielemann11}
{Thielemann}, F.-K., {Arcones}, A., {K{\"a}ppeli}, R., {et~al.} 2011, Progress
  in Particle and Nuclear Physics, 66, 346

\bibitem[{{Thielemann} {et~al.}(1990){Thielemann}, {Hashimoto}, \&
  {Nomoto}}]{Thielemann90}
{Thielemann}, F.-K., {Hashimoto}, M.-A., \& {Nomoto}, K. 1990, \apj, 349, 222

\bibitem[{{Timmes} {et~al.}(2003){Timmes}, {Brown}, \& {Truran}}]{Timmes03}
{Timmes}, F.~X., {Brown}, E.~F., \& {Truran}, J.~W. 2003, \apjl, 590, L83

\bibitem[{{Timmes} {et~al.}(1995){Timmes}, {Woosley}, \& {Weaver}}]{Timmes95}
{Timmes}, F.~X., {Woosley}, S.~E., \& {Weaver}, T.~A. 1995, \apjs, 98, 617

\bibitem[{{Tinsley}(1979)}]{Tinsley79}
{Tinsley}, B.~M. 1979, \apj, 229, 1046

\bibitem[{{Travaglio} {et~al.}(2015){Travaglio}, {Gallino}, {Rauscher},
  {R{\"o}pke}, \& {Hillebrandt}}]{Travaglio15}
{Travaglio}, C., {Gallino}, R., {Rauscher}, T., {R{\"o}pke}, F.~K., \&
  {Hillebrandt}, W. 2015, \apj, 799, 54

\bibitem[{{Travaglio} {et~al.}(2005){Travaglio}, {Hillebrandt}, \&
  {Reinecke}}]{Travaglio05}
{Travaglio}, C., {Hillebrandt}, W., \& {Reinecke}, M. 2005, \aap, 443, 1007

\bibitem[{{Tsujimoto} \& {Shigeyama}(2012)}]{Tsujimoto12}
{Tsujimoto}, T. \& {Shigeyama}, T. 2012, \apjl, 760, L38

\bibitem[{{Tucci Maia} {et~al.}(2015){Tucci Maia}, {Mel{\'e}ndez}, {Castro},
  {Asplund}, {Ram{\'{\i}}rez}, {Monroe}, {do Nascimento}, \&
  {Yong}}]{TucciMaia15}
{Tucci Maia}, M., {Mel{\'e}ndez}, J., {Castro}, M., {et~al.} 2015, \aap, 576,
  L10

\bibitem[{{Tucci Maia} {et~al.}(2014){Tucci Maia}, {Mel{\'e}ndez}, \&
  {Ram{\'{\i}}rez}}]{TucciMaia14}
{Tucci Maia}, M., {Mel{\'e}ndez}, J., \& {Ram{\'{\i}}rez}, I. 2014, \apjl, 790,
  L25

\bibitem[{{Tucci Maia} {et~al.}(2016){Tucci Maia}, {Ram{\'{\i}}rez},
  {Mel{\'e}ndez}, {Bedell}, {Bean}, \& {Asplund}}]{TucciMaia16}
{Tucci Maia}, M., {Ram{\'{\i}}rez}, I., {Mel{\'e}ndez}, J., {et~al.} 2016,
  \aap, 590, A32

\bibitem[{{van Kerkwijk} {et~al.}(2010){van Kerkwijk}, {Chang}, \&
  {Justham}}]{vanKerkwijk10}
{van Kerkwijk}, M.~H., {Chang}, P., \& {Justham}, S. 2010, \apjl, 722, L157

\bibitem[{{van Leeuwen}(2007)}]{vanLeeuwen07}
{van Leeuwen}, F. 2007, \aap, 474, 653

\bibitem[{{Vangioni} {et~al.}(2016){Vangioni}, {Goriely}, {Daigne}, {Fran{\c
  c}ois}, \& {Belczynski}}]{Vangioni16}
{Vangioni}, E., {Goriely}, S., {Daigne}, F., {Fran{\c c}ois}, P., \&
  {Belczynski}, K. 2016, \mnras, 455, 17

\bibitem[{{Venn} {et~al.}(2004){Venn}, {Irwin}, {Shetrone}, {Tout}, {Hill}, \&
  {Tolstoy}}]{Venn04}
{Venn}, K.~A., {Irwin}, M., {Shetrone}, M.~D., {et~al.} 2004, \aj, 128, 1177

\bibitem[{{Vincenzo} {et~al.}(2016){Vincenzo}, {Matteucci}, {Belfiore}, \&
  {Maiolino}}]{Vincenzo16}
{Vincenzo}, F., {Matteucci}, F., {Belfiore}, F., \& {Maiolino}, R. 2016,
  \mnras, 455, 4183

\bibitem[{{Wanajo} {et~al.}(2014){Wanajo}, {Sekiguchi}, {Nishimura}, {Kiuchi},
  {Kyutoku}, \& {Shibata}}]{Wanajo14}
{Wanajo}, S., {Sekiguchi}, Y., {Nishimura}, N., {et~al.} 2014, \apjl, 789, L39

\bibitem[{{Wang} {et~al.}(2013){Wang}, {Justham}, \& {Han}}]{Wang13}
{Wang}, B., {Justham}, S., \& {Han}, Z. 2013, \aap, 559, A94

\bibitem[{{Webbink}(1984)}]{Webbink84}
{Webbink}, R.~F. 1984, \apj, 277, 355

\bibitem[{{Winteler} {et~al.}(2012){Winteler}, {K{\"a}ppeli}, {Perego},
  {Arcones}, {Vasset}, {Nishimura}, {Liebend{\"o}rfer}, \&
  {Thielemann}}]{Winteler12}
{Winteler}, C., {K{\"a}ppeli}, R., {Perego}, A., {et~al.} 2012, \apjl, 750, L22

\bibitem[{{Yana Galarza} {et~al.}(2016){Yana Galarza}, {Mel{\'e}ndez},
  {Ram{\'{\i}}rez}, {Karakas}, {Asplund}, \& {Liu}}]{YanaGalarza16a}
{Yana Galarza}, J., {Mel{\'e}ndez}, J., {Ram{\'{\i}}rez}, I., {et~al.} 2016,
  ArXiv e-prints [\eprint[arXiv]{1602.08008}]

\bibitem[{{Yi} {et~al.}(2001){Yi}, {Demarque}, {Kim}, {Lee}, {Ree}, {Lejeune},
  \& {Barnes}}]{Yi01}
{Yi}, S., {Demarque}, P., {Kim}, Y.-C., {et~al.} 2001, \apjs, 136, 417

\bibitem[{{Yoshii} {et~al.}(1996){Yoshii}, {Tsujimoto}, \& {Nomoto}}]{Yoshii96}
{Yoshii}, Y., {Tsujimoto}, T., \& {Nomoto}, K. 1996, \apj, 462, 266

\bibitem[{{Yungelson}(2010)}]{Yungelson10}
{Yungelson}, L.~R. 2010, Astronomy Letters, 36, 780

\bibitem[{{Yungelson} \& {Livio}(2000)}]{Yungelson00}
{Yungelson}, L.~R. \& {Livio}, M. 2000, \apj, 528, 108

\end{thebibliography}

\begin{table*}
\caption{The [X/Fe] ratios from our analysis}\label{XFe_ratios}
\begin{center}
\begin{threeparttable}
\small
\begin{tabular}{l|cccccc} 
\hline\hline             
Star & C & O & Na & Mg & Al & Si \\ \hline
18Sco & $-$0.063$\pm$0.015 (5) & $-$0.050$\pm$0.009 (3) & $-$0.031$\pm$0.007 (4) & $-$0.014$\pm$0.012 (4) & $-$0.021$\pm$0.013 (6) & $-$0.007$\pm$0.009 (14) \\
HIP56948 & $-$0.011$\pm$0.015 (5) & $-$0.005$\pm$0.011 (3) & $-$0.004$\pm$0.011 (4) & $-$0.007$\pm$0.005 (3) & $-$0.008$\pm$0.005 (2) & 0.003$\pm$0.010 (12) \\
HIP102152 & 0.008$\pm$0.011 (4) & 0.009$\pm$0.008 (3) & $-$0.046$\pm$0.010 (3) & 0.025$\pm$0.016 (3) & 0.015$\pm$0.010 (6) & 0.002$\pm$0.007 (13) \\
HD20630 & $-$0.106$\pm$0.022 (4) & 0.008$\pm$0.019 (3) & $-$0.075$\pm$0.023 (3) & $-$0.083$\pm$0.017 (3) & $-$0.051$\pm$0.022 (6) & $-$0.027$\pm$0.017 (13) \\
HD202628 & $-$0.087$\pm$0.030 (4) & $-$0.044$\pm$0.009 (3) & $-$0.077$\pm$0.009 (3) & $-$0.020$\pm$0.018 (3) & $-$0.058$\pm$0.022 (6) & $-$0.010$\pm$0.009 (13) \\
HIP30502 & $-$0.065$\pm$0.010 (4) & $-$0.015$\pm$0.011 (3) & $-$0.036$\pm$0.005 (3) & 0.030$\pm$0.028 (3) & $-$0.001$\pm$0.024 (6) & 0.003$\pm$0.012 (13) \\
HIP73815 & $-$0.012$\pm$0.031 (4) & $-$0.008$\pm$0.012 (3) & $-$0.003$\pm$0.005 (3) & 0.021$\pm$0.025 (3) & $-$0.004$\pm$0.026 (6) & 0.007$\pm$0.011 (13) \\
HIP77883 & $-$0.015$\pm$0.012 (4) & 0.008$\pm$0.009 (3) & $-$0.019$\pm$0.010 (3) & 0.011$\pm$0.009 (3) & 0.006$\pm$0.012 (6) & 0.004$\pm$0.009 (13) \\
HIP89650 & 0.041$\pm$0.005 (4) & 0.045$\pm$0.025 (3) & 0.021$\pm$0.017 (3) & 0.036$\pm$0.018 (3) & 0.003$\pm$0.020 (6) & 0.010$\pm$0.014 (13) \\ \hline \hline
Star & S & K & Ca & Sc & Ti & V \\ \hline
18Sco & $-$0.038$\pm$0.016 (6) & $-$0.016$\pm$0.006 (1) & 0.002$\pm$0.009 (10) & $-$0.006$\pm$0.009 (11) & $-$0.002$\pm$0.010 (31) & $-$0.013$\pm$0.009 (9) \\
HIP56948 & $-$0.013$\pm$0.010 (4) & $-$0.011$\pm$0.006 (1) & 0.003$\pm$0.011 (11) & 0.003$\pm$0.009 (12) & 0.001$\pm$0.008 (20) & 0.015$\pm$0.013 (12) \\
HIP102152 & $-$0.001$\pm$0.026 (6) & 0.010$\pm$0.007 (1) & 0.010$\pm$0.008 (10) & $-$0.003$\pm$0.009 (9) & 0.012$\pm$0.009 (21) & $-$0.021$\pm$0.004 (9) \\
HD20630 & $-$0.026$\pm$0.016 (6) & 0.013$\pm$0.011 (1) & 0.020$\pm$0.018 (10) & $-$0.038$\pm$0.014 (9) & $-$0.022$\pm$0.013 (21) & $-$0.012$\pm$0.017 (9) \\
HD202628 & $-$0.053$\pm$0.034 (6) & 0.034$\pm$0.010 (1) & 0.030$\pm$0.013 (10) & 0.002$\pm$0.013 (9) & 0.015$\pm$0.009 (21) & $-$0.005$\pm$0.015 (9) \\
HIP30502 & $-$0.059$\pm$0.019 (6) & $-$0.019$\pm$0.007 (1) & 0.007$\pm$0.009 (10) & 0.017$\pm$0.013 (9) & 0.024$\pm$0.010 (21) & 0.005$\pm$0.009 (9) \\
HIP73815 & 0.012$\pm$0.025 (6) & 0.039$\pm$0.006 (1) & $-$0.004$\pm$0.015 (10) & 0.025$\pm$0.005 (9) & 0.005$\pm$0.011 (21) & $-$0.006$\pm$0.010 (9) \\
HIP77883 & 0.005$\pm$0.032 (6) & $-$0.010$\pm$0.005 (1) & 0.004$\pm$0.010 (10) & 0.009$\pm$0.008 (9) & 0.020$\pm$0.010 (21) & $-$0.002$\pm$0.011 (9) \\
HIP89650 & 0.015$\pm$0.020 (6) & 0.038$\pm$0.007 (1) & $-$0.002$\pm$0.014 (10) & 0.028$\pm$0.008 (9) & $-$0.005$\pm$0.010 (21) & $-$0.001$\pm$0.014 (9) \\ \hline \hline
Star & Cr & Mn & Co & Ni  & Cu & Zn \\ \hline
18Sco & 0.007$\pm$0.005 (21) & $-$0.011$\pm$0.005 (5) & $-$0.028$\pm$0.009 (9) & $-$0.013$\pm$0.010 (19) & $-$0.023$\pm$0.008 (4) & $-$0.034$\pm$0.004 (3) \\
HIP56948 & $-$0.002$\pm$0.009 (12) & 0.000$\pm$0.011 (6) & 0.008$\pm$0.009 (7) & 0.005$\pm$0.007 (13) & $-$0.005$\pm$0.010 (4) & 0.001$\pm$0.006 (3) \\
HIP102152 & $-$0.004$\pm$0.007 (16) & $-$0.029$\pm$0.009 (5) & $-$0.035$\pm$0.009 (8) & $-$0.024$\pm$0.010 (17) & $-$0.002$\pm$0.017 (4) & 0.010$\pm$0.014 (2) \\
HD20630 & 0.017$\pm$0.015 (16) & $-$0.015$\pm$0.020 (5) & $-$0.064$\pm$0.023 (9) & $-$0.057$\pm$0.018 (19) & $-$0.094$\pm$0.030 (4) & $-$0.068$\pm$0.009 (2) \\
HD202628 & 0.006$\pm$0.014 (16) & $-$0.052$\pm$0.010 (5) & $-$0.044$\pm$0.020 (9) & $-$0.060$\pm$0.014 (19) & $-$0.096$\pm$0.037 (4) & $-$0.056$\pm$0.006 (2) \\
HIP30502 & $-$0.003$\pm$0.007 (16) & $-$0.033$\pm$0.006 (5) & 0.002$\pm$0.009 (9) & $-$0.013$\pm$0.009 (19) & $-$0.013$\pm$0.005 (4) & $-$0.042$\pm$0.012 (2) \\
HIP73815 & $-$0.008$\pm$0.009 (16) & $-$0.013$\pm$0.005 (5) & 0.000$\pm$0.011 (9) & $-$0.015$\pm$0.010 (19) & 0.032$\pm$0.036 (4) & 0.019$\pm$0.015 (2) \\
HIP77883 & 0.005$\pm$0.009 (16) & $-$0.004$\pm$0.009 (5) & 0.003$\pm$0.022 (9) & $-$0.007$\pm$0.021 (19) & 0.016$\pm$0.024 (4) & 0.016$\pm$0.004 (2) \\
HIP89650 & $-$0.004$\pm$0.010 (16) & $-$0.005$\pm$0.013 (5) & 0.021$\pm$0.014 (9) & 0.004$\pm$0.011 (19) & 0.030$\pm$0.024 (4) & 0.021$\pm$0.006 (2) \\ \hline \hline
Star & Y & Ba & La & Ce  & Nd & Eu \\ \hline
18Sco & 0.053$\pm$0.010 (5) & 0.066$\pm$0.009 (3) & 0.084$\pm$0.010 (3) & 0.066$\pm$0.015 (8) & 0.103$\pm$0.020 (6) & 0.138$\pm$0.051 (4) \\
HIP56948 & 0.008$\pm$0.007 (5) & 0.002$\pm$0.005 (3) & 0.006$\pm$0.013 (2) & 0.027$\pm$0.027 (4) & 0.033$\pm$0.009 (1) & 0.057$\pm$0.015 (1) \\
HIP102152 & $-$0.120$\pm$0.036 (5) & $-$0.014$\pm$0.004 (2) & $-$0.029$\pm$0.035 (1) & $-$0.018$\pm$0.009 (1) & 0.013$\pm$0.017 (3) & 0.053$\pm$0.034 (1) \\
HD20630 & 0.088$\pm$0.065 (5) & 0.192$\pm$0.015 (3) & 0.170$\pm$0.042 (1) & 0.143$\pm$0.016 (1)& 0.095$\pm$0.026 (2) & 0.085$\pm$0.008 (1) \\
HD202628 & 0.068$\pm$0.075 (5) & 0.206$\pm$0.021 (3) & 0.196$\pm$0.024 (1) & 0.182$\pm$0.009 (1)& 0.176$\pm$0.005 (2) & 0.060$\pm$0.010 (1) \\
HIP30502 & $-$0.035$\pm$0.044 (5) & 0.037$\pm$0.016 (3) & 0.010$\pm$0.006 (1) & 0.089$\pm$0.003 (1)& 0.086$\pm$0.031 (2) & 0.032$\pm$0.026 (1) \\
HIP73815 & $-$0.124$\pm$0.041 (5) & $-$0.020$\pm$0.033 (3) & $-$0.049$\pm$0.017 (1) & $-$0.019$\pm$0.009 (1) & $-$0.003$\pm$0.036 (2) & $-$0.008$\pm$0.011 (1) \\
HIP77883 & $-$0.097$\pm$0.030 (5) & $-$0.016$\pm$0.014 (3) & $-$0.011$\pm$0.009 (1) & $-$0.042$\pm$0.007 (1) & $-$0.022$\pm$0.031 (2) & $-$0.002$\pm$0.008 (1) \\
HIP89650 & $-$0.050$\pm$0.033 (5) & $-$0.004$\pm$0.011 (3) & 0.019$\pm$0.015 (1) & 0.043$\pm$0.009 (1) & 0.012$\pm$0.054 (2) & 0.021$\pm$0.009 (1) \\ \hline \hline
\end{tabular}
\begin{tablenotes}
      \tiny
      \item In brackets we report the number of lines that have been detected for each element.
    \end{tablenotes}
\end{threeparttable}
\end{center}
\end{table*}




\end{document}